\documentstyle[12pt]{article}

\def\gsim{\mbox{$\stackrel{_>}{_\sim}$}} %Maggiore o circa%
\def\lsim{\mbox{$\stackrel{_<}{_\sim}$}} %Minore   o circa%

\def\bec{\begin{center}}
\def\eec{\end{center}}

\def\beq{\begin{equation}}
\def\eeq{\end{equation}}

\renewcommand{\frac}[2]{{{\displaystyle #1}\over{\displaystyle #2}}}

\def\MeV{\mbox{ MeV}}

\def\BeSv{\mbox{$^{7}$Be}}     % Isotopes
\def\BHt{\mbox{$^{8}$B}}

\def\calA{\mbox{$\cal A$}}     % Calligraphic letters

\def\calN{\mbox{$\cal N$}}

\def\calR{\mbox{$\cal R$}}
\def\calS{\mbox{$\cal S$}}

\def\dms{\mbox{$\Delta m^2$}}     % neutrino MSW parameters

\def\SdTvS{\mbox{$\sin^2 2\theta_V$}}

\def\numt{\mbox{$\nu_{\mu(\tau)}$}} %  nu_mu(tau) neutrinos
\def\nue{\mbox{$\nu_e$}}            %  nu_e
\def\Ye{\mbox{$Y_e$}}           % Isotopic ratio

\def\Enu{\mbox{$E_\nu$}}         % Neutrino Energy
\def\Te{\mbox{$T_e$}}            % Electron Kinetic energy
\def\TeTh{\mbox{$T_{e,th}$}}     % Threshold Electron energy
\def\SeZr{\mbox{$\calS_{0}$}}       %    Undistorted 
\def\Ses{\mbox{$\calS^s$}}          %       With s index
\def\SeD{\mbox{$\calS^D$}}          %       Day
\def\SeN{\mbox{$\calS^N$}}          %       Night
\def\SeC{\mbox{$\calS^C$}}          %       Core
\def\SeM{\mbox{$\calS^D$}}          %       Mantle
\def\Res{\mbox{$\calR^s$}}          %       "     with s index
\def\ReZr{\mbox{$\calR_{0}$}}       %   Solar Standard Model
\def\ReZrs{\mbox{$\calR_0^s$}}      %        "   "       "   with s index
\def\ReZrD{\mbox{$\calR_{0}^D$}}    %        "   "       "   for Day
\def\ReZrN{\mbox{$\calR_{0}^N$}}    %        "   "       "   for Night
\def\ReZrC{\mbox{$\calR_{0}^C$}}    %        "   "       "   for Core
\def\ReZrM{\mbox{$\calR_{0}^M$}}    %        "   "       "   for Mantle
\def\ReD{\mbox{$\calR^D$}}          %     Day
\def\AsymSs{\mbox{$\calA_{D-N}^s$}} %      "     with s index
\def\AsymSC{\mbox{$\calA_{D-N}^C$}} %      core
\def\AsymRs{\mbox{$A_{D-N}^s$}}     %        "      with s index
\def\AsymRN{\mbox{$A_{D-N}^N$}}     %      night
\def\AsymRC{\mbox{$A_{D-N}^C$}}     %      core
\def\AsymRM{\mbox{$A_{D-N}^M$}}     %      mantle
\def\AsymRNCM{\mbox{$A_{D-N}^{N,C,M}$}}   %      night - core - mantle
\def\AsymRNM{\mbox{$A_{D-N}^{N(M)}$}}   %      night - core - mantle
\def\deltaSs{\mbox{$\calN^s$}}   %   Generical
\def\deltaSD{\mbox{$\calN^D$}}   %   Day
\def\PTot{\mbox{$P_{\oplus}$}}          % Total probability
\def\PTots{\mbox{$P^s_\oplus$}}         %     "             for sample s
\def\DAY{\mbox{\em{Day}}}                        %    Day
\def\night{\mbox{\em{Night}}}                    %    Night
\def\core{\mbox{\em{Core}}}                      %    Core
\def\mantle{\mbox{\em{Mantle}}}                  %    Mantle
\def\dseedEe{\mbox{$\frac{d\, \sigma_{\nu_e  } (\Te,E_\nu)}{d\,\Te}$}}
\def\dsemdEe{\mbox{$\frac{d\, \sigma_{\nu_\mu} (\Te,E_\nu)}{d\,\Te}$}}
\def\daynight{D-N}                        % Day/Night expresion
\def\SK{Super - Kamiokande}

\def\electron{\mbox{$e^-$}}

\def\FigSKSpectra{1}   % Replaces \ref{fig:sk:spectra}
\def\FigContour{3}

\def\TabEventRates{I}  
\def\TabAsymmetries{II}  
\def\TabAsymmetriesMeV{III}

\begin{document}
\sloppy

\vspace{0.5cm}
%
% References Pointer
%
{\normalsize
\begin{flushright}
\begin{tabular}{l}
SISSA 17/97/EP\\
arch-iv/9705392\\
May, 1997\\

\end{tabular}
\end{flushright}
}
\vspace{1cm}

%%%%%%%%%%%%%
% The Title %
%%%%%%%%%%%%%
\begin{center}
{\Large
A Study of the Day - Night Effect for the \SK\  Detector: II.
% \vspace{0.25cm}
Electron Spectrum Deformations and Day - Night Asymmetries
}
\end{center}

%%Authors
\begin{center}
M. Maris $^{\mbox{a,b)}}$ and 
S. T. Petcov 
\footnote{
Also at: Institute of Nuclear
Research and Nuclear Energy, Bulgarian Academy of Sciences,
1784 Sofia, Bulgaria.} $^{\mbox{a,c)}}$ \\
$^{\mbox{a)}}$ Scuola Internazionale Superiore di Studi Avanzati,
Trieste, Italy.\\
$^{\mbox{b)}}$ INFN - Sezione di Pavia, Pavia, Italy.\\
$^{\mbox{c)}}$ INFN - Sezione di Trieste, Trieste, Italy.
\end{center}

\bec
\abstract{
\noindent
Using the results of a high precision calculation of the solar neutrino 
survival probability for Earth crossing neutrinos in the case of MSW 
$\nu_e \rightarrow \nu_{\mu(\tau)}$\ transition solution of the solar 
neutrino problem, performed in an earlier 
study, we derive predictions for the one-year averaged day-night (D-N) 
asymmetry in the deformations of the $e^-$ - spectrum to be measured with 
the Super - Kamiokande detector, and for the D-N asymmetry in the energy- 
integrated one year signal in this detector.
The asymmetries are calculated for solar $\nu_e$\ crossing the Earth 
mantle only, the core and the (mantle + core) for a large representative 
set of values of the MSW transition parameters \dms\ and 
\SdTvS\ from the ``conservative'' MSW solution region 
obtained by taking into account possible uncertainties in the values of 
the $^8$B and $^7$Be neutrino fluxes. The effect of the uncertainties
in the value of the bulk matter
density and in the chemical composition of the core, 
on the D-N asymmetry predictions is discussed. 
It is shown, in particular, that for $\SdTvS \leq 0.013$\ 
the one year
average D-N asymmetry for neutrinos crossing the Earth core can be 
larger than the asymmetry for (only mantle crossing + core crossing) 
neutrinos by a factor of up to six.
Iso - (D-N) asymmetry contours in the $\dms - \SdTvS$\ 
plane for the Super - Kamiokande detector are derived in the 
region $\SdTvS \gsim 10^{-4}$\ for only mantle crossing, core 
crossing and (only mantle crossing + core crossing) neutrinos.
Our results indicate that the Super - Kamiokande experiment might be able 
to test the $\SdTvS \le 0.01$\ region of the MSW solution of 
the solar neutrino problem by performing selective D-N asymmetry measurements.

\vspace{3mm}
 }
\eec

\newpage

\section{Introduction}
Assuming that the solar neutrinos undergo two - neutrino MSW 
$\nue \rightarrow \numt$\ 
transitions in the Sun, and that these transitions are at the origin of the 
solar neutrino deficit, we have performed in 
\cite{ArticleI}\ a high - 
precision calculation of the one - year averaged solar \nue\ survival 
probability for Earth crossing neutrinos, 
$\PTot(\nue\rightarrow\nue)$, reaching the \SK\ detector.
The probability $\PTot(\nue\rightarrow\nue)$\ 
was calculated by using, in particular, 
the elliptical orbit approximation (EOA) 
to describe the movement of the Earth around the Sun. 
Results for $\PTot(\nue\rightarrow\nue)$\ as a function 
of $\Enu/\dms$, $\Enu$\ and \dms\ 
being the neutrino energy and the neutrino mass 
squared difference, have been 
obtained for neutrinos crossing the Earth mantle only, 
the core, the inner 2/3 of 
the core and the mantle + core (full night) for a 
large representative set of values of 
\SdTvS, where $\theta_V$\ is the neutrino mixing angle in vacuum, from the 
''conservative'' MSW solution region in the 
\dms\ - \SdTvS\ plane, derived by taking 
into account the possible uncertainties in the fluxes 
of \BHt\ and \BeSv\ neutrinos (see, e.g., ref. \cite{SPnu96}; for earlier
studies see ref. \cite{PKAS94}).

We have found in \cite{ArticleI}, in particular, that for 
$\SdTvS \leq 0.013$\ the one - year averaged \daynight\ asymmetry
\footnote{A rather complete list of references on the \daynight\
effect is given in ref. \cite{ArticleI}.}
in the 
probability $\PTot(\nue\rightarrow\nue)$\ for neutrinos crossing the 
Earth core can be larger than the asymmetry in the probability for 
(only mantle crossing + core crossing) 
neutrinos by a factor of up to six.
The enhancement is even larger for neutrinos crossing the inner 2/3 of 
the core. We have also pointed out to certain subtleties in the calculation 
of the time averaged \nue\ survival probability 
$\PTot(\nue\rightarrow\nue)$\
for neutrinos crossing the Earth, which 
become especially important when $\PTot(\nue\rightarrow\nue)$\ is 
computed, for instance, for the core crossing neutrinos only
\footnote{For further details concerning the technical 
aspects of the calculations see 
ref. \cite{ArticleI} as well as ref. \cite{Maris:1997}.}. 

In the present article we use the results obtained in \cite{ArticleI}\ to
investigate the \daynight\ asymmetries in the spectrum of the recoil 
electrons from the reaction $\nue + e^- \rightarrow \nue + e^-$\ caused 
by the \BHt\ neutrinos and in the energy-integrated event rate, 
to be measured by the \SK\ experiment. 
If the solar \BHt\ neutrinos take part in MSW transitions, the \BHt\ 
neutrino spectrum will be deformed by the MSW effect in the Sun, and, for 
the neutrinos crossing the Earth, by the MSW effect in the Earth.
This creates a difference between the (deformed) spectra of neutrinos 
detected during the day and during the night, which is reflected in the 
corresponding day and night recoil - $e^-$\ spectra. 
We have computed in \cite{SKDNII:spectrum}\ the \daynight\ asymmetry in 
the recoil-$e^-$\ spectrum for the 
same large set of representative values of \dms\ and \SdTvS\ from the 
``conservative'' MSW solution region for which the results in 
\cite{ArticleI}\ have been obtained. The \daynight\ asymmetry in the 
$e^-$-spectrum is found for neutrinos crossing the Earth mantle only, 
the core  and the mantle + core.
Here we have included only few representative plots showing the 
magnitude of the \daynight\ asymmetry in the recoil-e$^{-}$ spectrum 
to be expected in the case of the 
two - neutrino MSW $\nue \rightarrow \numt$\ 
transition solution of the solar neutrino problem.
The spectrum asymmetry for the sample of events due to core crossing 
neutrinos only is strongly enhanced for 
$\SdTvS~\lsim~0.013$\ with respect to the analogous asymmetries for the 
mantle and for the (only mantle crossing + core crossing) neutrinos. 
We present also detailed results for the one-year averaged \daynight\ 
asymmetry in the \SK\ signal for the indicated three samples of events.
We find that indeed for $\SdTvS \leq 0.013$\ the asymmetry in the sample 
corresponding to core crossing neutrinos can be larger than the asymmetry 
in the sample produced by only mantle crossing or by 
(only mantle crossing + core crossing ) neutrinos by a factor of up to six.
The dependence of the D-N asymmetries in the three samples on the threshold
e$^{-}$ kinetic energy being used for event selection is also investigated.

We derive iso - (\daynight) asymmetry contours in the region of 
$\SdTvS~\gsim~10^{-4}$\ in the \dms - \SdTvS\ plane for the signals in 
the \SK\ detector produced by neutrinos crossing the mantle, the core 
and mantle + core. The iso-asymmetry contours for the sample of events due to
core crossing neutrinos are obtained for two values of the
fraction of electrons, $Y_e$, in the core: for $Y_e = 0.467$ and 0.500
\footnote{Iso - (D-N) asymmetry contour plots for the full night (i.e., mantle + 
core) signal 
for the \SK\ detector for one value of $Y_e$ have been obtained 
in refs. \cite{PKDN96, Lisi:Montanino:1997}.}.

Our results confirm the conclusion 
drawn in \cite{ArticleI}\ 
that the \SK\ detector might be able to test the $\SdTvS~\lsim~0.01$\ 
region of the MSW solution of the solar neutrino problem.

\section{\daynight\ Effect Related Observables}

In \cite{ArticleI}\ we have considered four possible groups or samples 
of solar neutrino events depending on their detection time.
We have labeled these samples as \DAY, \night, \core\ and \mantle, where
\DAY\ and \night\
 samples consist of the solar neutrino events detected respectively during 
the day and during the night, while the other two samples are 
formed by the events induced by the 
neutrinos which cross the Earth core (\core) and by the neutrinos
which does not cross it (\mantle).
All quantities analyzed in this paper (MSW probability, recoil-\electron\ 
spectrum, event rate, asymmetries) refer to one of these samples and 
correspondingly carry one of the indices $D$, $N$, $C$ and $M$, standing 
respectively for \DAY, \night, \core, and \mantle.
The recoil-\electron\ spectra associated with the four samples are
denoted by $\Ses(\Te)$, where $s = D$, $N$, $C$, $M$, and \Te\ is
the recoil-\electron\ kinetic energy, while for the event rates we will use 
the notation \Res.
The symbols $\SeZr(\Te)$\ and $\ReZrs$\ will be used to
denote the recoil-\electron\ spectrum and event rates for massless 
(``conventionally'' behaving) neutrinos, computed using 
the predictions of a given 
reference standard solar model.
Correspondingly, one has $\ReZrD = \ReZrN = \ReZrC = \ReZrM~ \equiv \ReZr$.
The spectra $\Ses(\Te)$, $\SeZr(\Te)$\ and the event rates $\ReZr$, 
\Res\ considered in the present article are one year averaged spectra and 
event rates.

In the present analysis we use the model of ref. \cite{BP95}\ 
with heavy elements diffusion as a reference solar model.
As is well known, the shape of the spectrum of \BHt\ neutrinos, and consequently
 the shape of the recoil \electron\ spectrum, is solar model independent.
The event rates \ReZr\ and \Res\ depend on the reference 
solar model prediction for 
the total flux of \BHt\ neutrinos.
However, the \daynight\ asymmetries we are going to consider, do not depend on 
the total \BHt\ neutrino flux and therefore are solar model 
independent quantities as well.

The spectra $\SeZr(\Te)$\ and $\Ses(\Te)$\ are given by the 
following standard expressions:
\bec\beq\label{eq:electron:spectra:SSM}
{\displaystyle
     \SeZr(\Te) = \Phi_B
     {\displaystyle \int_{ \Te\left(1 + 
            {{m_e}\over{2 T_e}}\right)} 
%^{14.4 \mbox{ MeV} } 
}
       d\Enu\, n(E_{\nu})\, \dseedEe
},
\eeq\eec

\noindent
and
\bec\beq
\begin{array}{lll}\label{eq:electron:spectra:msw}
    \Ses(\Te) =& \Phi_B {\displaystyle \int_{ \Te\left(1 +
                     {{m_e}\over{2 T_e}}\right)}
%^{\displaystyle +14.4 \mbox{ MeV}}
}
                     d\Enu\, 
                     nu(E_{\nu}) \,
 &\left[ \dseedEe P_\oplus^s\left( \nue \rightarrow \nue
%%%%{{\Enu}\over{\Delta m^2}}
\right) + \right.\\
         &\\
         && + \left.\dsemdEe 
              \left(1-P_\oplus^s\left( \nue \rightarrow \nue
%%%%{{\Enu}\over{\Delta m^2}}
\right)\right) 
    \right].\\ 
\end{array} 
\eeq\eec

\noindent
Here $\Phi_B$ is the total \BHt\ neutrino flux, 
\Enu\ is the incoming \BHt\ neutrino energy, $m_e$\ is the electron mass, 
$n(E_{\nu})$ is the normalized to one \BHt\ neutrino spectrum
\cite{Bahcall:etal:1996},
$\PTots(\nue \rightarrow \nue)$\ is the one year averaged solar \nue\ 
survival probability for \DAY, \night, \core, and \mantle\ samples, and
$d \, \sigma_{\nu_e(\nu_\mu)} (\Te,\Enu) / d\,\Te$\ is the
differential $\nue~(\numt) - \electron$\ elastic scattering cross section
\cite{Bahcall:Kamionkowski:Sirlin:1995}.
For the corresponding energy integrated event rates we have:
\bec\beq
\begin{array}{lll}
  \ReZr(\TeTh) &=&  {\displaystyle \int_{\TeTh}
%^{+14.1 \mbox{ MeV}} 
} 
d\Te\,\SeZr(\Te),\\
   &&\\
  \Res(\TeTh) &= & {\displaystyle \int_{\TeTh}
%^{+14.1 \mbox{ MeV}}
 } d\Te \, \Ses(\Te),\\
\end{array}
\eeq\eec

\noindent
where \TeTh\ is the recoil-\electron\ kinetic energy threshold of the \SK\ 
detector. The results reported in the present study are obtained for
$\TeTh = 5$\ MeV.

In this article we present results for three observables relevant to 
\daynight\ effect which can be measured with the \SK\ detector. 
The first is the distortion of the recoil-\electron\ spectrum due to the 
MSW effect for the four different event samples:
\bec\beq
\deltaSs(\Te) = \frac{\Ses(\Te)}{\SeZr(\Te)}, \mbox{\hspace{1cm}} 
            s = \mbox{$D$, $N$, $C$, $M$}.
\eeq\eec

\noindent
In the absence of the MSW effect, or in the case of energy - independent 
(constant) reduction of the \BHt\ \nue\ flux at $\Enu \ge 5$ MeV, we 
would have $\deltaSs(\Te)$\ = const. 
The spectrum ratio $\deltaSs(\Te)$\ shows the magnitude and the shape of the
\electron\ spectrum deformations (with respect to the standard spectrum)
due to the MSW effect taking place in the Sun only, as well as 
in the Sun and in the Earth mantle, core and mantle + core.

The second observable we consider is the \daynight\ asymmetry in the 
recoil-\electron\ spectrum for the three solar neutrino event samples, 
associated with the MSW effect in the Earth:
\bec\beq
     \AsymSs(\Te) = 2 \,\, \frac{\Ses(\Te)-\SeD(\Te)}{\Ses(\Te)+\SeD(\Te)},
      \mbox{\hspace{1cm}}
       s = \mbox{$N$, $C$, $M$}.
\eeq\eec

\noindent The third observable is the energy integrated event rate asymmetry:
\bec\beq
     \AsymRs (T_{e,th}) = 2 \,\, \frac{\Res - \ReD}{\Res+\ReD}, 
      \mbox{\hspace{1cm}}
       s= \mbox{$N$, $C$, $M$}.
\eeq\eec

\noindent
Both these observables are solar model independent.

The observation of a nonzero \daynight\ asymmetry $\AsymRs \neq 0$\ and/or
$\AsymSs(\Te) \neq 0$\ would be a very strong evidence (if not a proof) that 
solar neutrinos undergo MSW transitions.
If, on the other hand, no statistically significant \daynight\ effect 
would be found within the accuracy of the measurements to be reached in 
the \SK\ experiment, a large region of the MSW solution values of the 
parameters \dms\ and \SdTvS\ would be excluded.

The one year averaged spectrum ratio 
$\deltaSs(\Te)$\, $s= D$, $N$, $C$, $M$,
and the spectrum and event rate \daynight\ asymmetries $\AsymSs(\Te)$\ and 
$\AsymRs$, $s= N$, $C$, $M$, for the \SK\ detector have been calculated for
36 pairs of values of \dms\ and \SdTvS\ distributed evenly in the 
``conservative'' MSW solution region (see \cite{ArticleI, SPnu96}). 
These values together with the corresponding results for 
\AsymRNCM\ as well as for the ratio $\AsymRC / \AsymRN$,
are given in Table II.
We present here graphically in Figs. 2.1 - 2.12 only some of the results 
obtained for $\deltaSs(\Te)$\ and $\AsymSs(\Te)$\ 
(see ref. \cite{SKDNII:spectrum} for a more complete set of figures). 
Figs. 3a - 3c contain plots of contours in the $\dms$\ - $\SdTvS$\ plane,
corresponding to fixed values of the event rate asymmetries 
\AsymRNCM\ (iso - (\daynight) asymmetry contours) in the 
region of $\SdTvS \gsim 10^{-4}$.

\section{Earth Model Uncertainties}

The Earth electron number density $n_e(x)$, which enters in the 
calculations of the \daynight\ effect, is proportional to the product 
$Y_e(x) \rho(x)$,
where $Y_e(x)$\ is the number of electrons per nucleon and $\rho(x)$\ 
is the matter density at distance
$x$ from the center of the Earth. 
While both $Y_e(x)$\ and $\rho (x)$ are functions of the local 
chemical composition, $\rho(x)$\ is 
also a function of the phase (solid, liquid, crystal) of the state of
matter. The phase can be inferred
from measurements of the propagation of $s$ and $p$ seismic waves 
in the Earth. 
These measurements permitted to reconstruct the approximate 
onion-like structure of the
Earth's interior, each strata being characterized by materials 
which are in definite phases having definite chemical composition.

As already was mentioned,
the two main components of the Earth 
structure are the core and the mantle. 
Both of them are characterized by a strong differentiation in both
the chemical composition and the state phases.
The chemical composition of the core and of the mantle 
cannot be inferred from the seismic data only. It is deduced 
by comparing results from laboratory experiments, the seismic data, 
and the predictions of models
(for further details see, e.g., refs.
\cite{Stacey:1977,Jeanloz:1990,Dziewonski:1981,Allegre:etal:1995}).
In laboratory experiments one studies the chemical and physical 
behavior of the 
materials at the Earth interior conditions, where the temperature
and the pressure reach values of  
the order of few $\times 10^3$\ K  and  
few$\times 10^{11}$ Pa, to obtain a quantitative measure of the 
correlation between pressure, density, composition and the sound velocity.
The interior Earth conditions are  reproduced by studying samples
subjected to high pressures and high temperatures
in either a diamond-anvil-cell or in shock-wave experiments. 
One of the major aims of these experiments is to 
produce a realistic model for the Earth's
chemical composition.
Since it is supposed that the Earth formed as a result of
gravitational accretion 
of planetesimals which are believed to be also the source of meteorites, 
it is suggested that the chemical composition of the Earth interior
resembles to a large extent that of meteorites. 
The mantle is expected to have a composition similar to the composition of the 
{\em chondritic} meteorites consisting mainly of (Mg,Fe)SiO$_3$. The core
composition is supposed to be similar to that of the 
the iron meteorites, i.e., an alloy of Fe and Ni with at most 10\%\
(by weight) of Ni.
However, while the seismic constraints are compatible with the chemical 
composition of the mantle thus deduced, they are not compatible with 
the indicated core composition. The outer core
has to be an alloy of iron with 10\%\ of some lighter element such as
O, S, Si.
The presence of light elements in the outer core 
may be due to the drag of materials from the
mantle toward the Earth center driven by the gravitational settling of the 
molten iron during the inner Earth differentiation. 
It may also be due to chemical instabilities caused 
by strong chemical in-equilibrium
at the core/mantle interface, which can produce chemical reactions and/or 
diffusion enriching the core with light elements and the mantle with heavy 
ones. 

Calculations based on the various possible chemical 
compositions of the two major Earth substructures 
show that $Y_e \approx (0.49 - 0.50)$\ for the 
mantle, while for
the core $Y_e \approx (0.46 - 0.48)$.
At the same time the bulk matter density of the core $\rho$\ 
can be inferred from the seismological data with 
an uncertainty which is estimated to be between 5\% and $10\%$,
the latter value representing a conservative upper limit. 
Thus, the uncertainty in the knowledge
of $\rho$\ is the major source of uncertainty in the quantity
$Y_e \rho$\ for the core. 

 Other source of uncertainty in the Earth model 
are the thickness
of the core/mantle boundary region and the value of the core radius.
Both of them are strongly constrained by seismological data and
the corresponding uncertainties do not exceed $\sim 10~$km.
The uncertainties they introduce in the magnitude of the 
\daynight\ asymmetries are negligible
because the corresponding residence time associated with them is 
exceedingly small.  
For the same reason the detailed structure of the core-mantle boundary 
region, more specifically, the change of density in this region 
(continuous versus discontinuous) is not important for a calculation of the
\daynight\ effect with a precision of (1 - 2)\% \cite{Bertotti:Maris:1996}.

\section{\daynight\ Effect and the Recoil-\electron\ Spectrum}

 In Figs. 1 and 2.1 - 2.12 we show examples respectively 
of the predicted recoil - e$^{-}$
spectrum, $\SeZr(\Te)$, the spectrum distortions due to the 
MSW effect in the Earth
and/or in the Sun, $\Ses(\Te)$, and of the
ratio of spectra, $\deltaSs(\Te)$, and the
\daynight\ asymmetries in the spectrum for the
different event samples, $\AsymSs(\Te)$. This is done for  
representative subset of neutrino parameters \dms\ and \SdTvS,
chosen from the set listed in Table \TabEventRates.
% Since the Earth effect in the recoil-e$^{-}$ spectrum is in general not large,
% as Fig. 1 indicates, its
The ``presence'' of the Earth effect in the 
e$^{-}$-spectrum is better illustrated by the ratio
$\deltaSs(\Te)$\ and by the 
asymmetries $\AsymSs(\Te)$\ than by the spectra $\Ses(\Te)$.
For this reason we report as an example in Fig. 1 only one set of spectra 
$\Ses(\Te)$\ calculated for 
$\SdTvS = 0.01$, $\dms = 5 \times 10^{-5}$\ eV$^2$\ and $Y_e = 0.467$. 
The {\em upper solid} line in Fig. 1 is the standard e$^{-}$-spectrum
$\SeZr(\Te)$, while 
$\SeD(\Te)$, $\SeN(\Te)$, $\SeC(\Te)$\ and $\SeM(\Te)$\ are represented 
respectively by the {\em long - dashed}, the {\em short - dashed}, 
the {\em lower solid} and the {\em dotted} lines.
It should be noted that since in the \SK\ experiment 
the background rejection is achieved, in particular, 
through a cut in the angle between the scattered electron momentum
direction and the Sun's
direction, the measured spectra (and their related distortions and \daynight\ 
asymmetries) depend, in general, on the cut being used.
For the \SK\ detector e$^{-}$ kinetic energy threshold of 5 MeV, however,
the effect of the angular cut on 
the spectra of interest, as can be shown, is negligible
\footnote{The effect of the indicated angular cut on the spectra of interest
becomes important for $\Te ~\lsim~ 4~$MeV.}.

The e$^{-}$-spectrum distortions $\deltaSs(\Te)$ 
and \daynight\ asymmetries $\AsymSs(\Te)$\
depicted in Figs. 2.1 - 2.12 (the upper and the lower frame of each
figure) were computed for $Y_e = 0.467$.
The enhancement of the spectrum distortions and of the \daynight\
asymmetry in the \core\ sample is clearly seen,
especially for $\SdTvS \lsim 0.013$.
The magnitude of the Earth effect in the spectrum 
is sensitive to the value of \dms, since spectral 
signatures such as kinks, knees and peaks 
(that may be associated with similar features in the   
the corresponding solar $\nu_e$ survival probabilities
reported in ref. \cite{ArticleI}) change their position
in the kinetic energy window when \dms\ changes 
(see Figs. 2.1, 2.2 and 2.3 as well as Figs.
2.4, 2.5 and 2.6).

 Both the spectra $\Ses(\Te)$, the spectral distortions 
$\deltaSs(\Te)$ shown in Figs. 1 and 2.1 - 2.12
and the event rates given in 
Table \TabEventRates\ are normalized to the event rate predicted for standard
neutrinos in the reference solar model \cite{BP95}.
That means that if the neutrino parameters belong to the  
``conservative'' MSW solution regions but lie outside 
the solution regions obtained within the reference solar model,
one should re-scale $\Ses(\Te)$, $\deltaSs(\Te)$ and the event rates 
in Table I, multiplying them by the factor $f_B = \Phi^{MSW}_{B}/\Phi_{B}$,
where $\Phi^{MSW}_{B}$ is the $^{8}$B neutrino flux which was used in the MSW 
analysis of the solar neutrino data and for which the corresponding solution
region was obtained (see refs. \cite{SPnu96, PKAS94} for details).
We give in Table I the values of the factor $f_B$ associated with each pair 
of values of \dms\ and \SdTvS, for which the event rates listed in Table I
and the quantities $\Ses(\Te)$\ and 
$\deltaSs(\Te)$\ depicted in Figs. 1 and 2.1 - 2.12 have been calculated. 
 
  The differences between the \DAY\ and the \night\ (\mantle) sample
e$^{-}$ spectra are hardly observable, as Figs. 2.1 - 2.12 indicate.
However, as it follows from Figs. 2.4, 2.5, 2.7, 2.10 and 2.12,
the \core\ sample spectrum can differ significantly 
(both in shape and magnitude) from the \DAY\ sample spectrum. 
For $\SdTvS \geq 0.008$ and for a large number of values of \dms\ 
from the ``conservative'' MSW solution region the \core\ sample
D-N asymmetry in the spectrum $\AsymSC(\Te)$\ is bigger than 10\% 
at least in a significant subinterval of values of $\Te$
from the relevant interval (5 - 14) MeV. In certain cases
the asymmetry $\AsymSC(\Te)$\ can reach values of 20\% - 30\% (Figs. 2.5, 2.6 
and 2.8) and even 50\% (Fig. 2.10). At the same time, for certain 
pairs of values of  \dms\ and \SdTvS, $\AsymSC(\Te)$\ is rather small in
the whole e$^{-}$ kinetic energy interval of interest. For instance,
if $\dms = 4\times 10^{-5}~\mbox{eV}^2$ ($5\times 10^{-5}~\mbox{eV}^2$)
and $\SdTvS = 0.3~(0.7)$ we have $\AsymSC(\Te) \leq 6\%~(7.5\%)$.
In any case, the measurement of the \core\ sample e$^{-}-$spectrum as 
well as of the asymmetry $\AsymSC(\Te)$\ can allow to obtain additional
constraints on the MSW transition parameters \dms\ and \SdTvS. 

 Finally, we would like to note that,  
we have performed a comparison between the spectra and the 
related spectrum distortions and \daynight\ asymmetries, 
computed for $Y_e=0.467$\ and for $Y_e = 0.50$.
The differences are relatively small if one takes into account
the precision in the measurement of the recoil-e$^{-}$ spectrum,
which is expected to be reached in the \SK\ experiment. 
The ratio $\deltaSs(\Te)$ and the \daynight\ asymmetries
$\AsymSs(\Te)$\ computed for $Y_e = 0.50$\ and the set of neutrino parameters
\SdTvS\ and \dms\ given in Table I
can be found in ref. \cite{SKDNII:spectrum}.

\section{Energy-Integrated \daynight\ Asymmetries}
Figures \FigContour a, \FigContour b and \FigContour c
represent the iso-(\daynight) asymmetry contour plots for the 
\night, \core\ and \mantle\ samples. 
Shown are also the ``conservative'' regions of the two MSW solutions
of the solar neutrino problem (dashed lines) as well as 
the solution regions obtained in the reference solar model
\cite{BP95} (dotted lines). 
The iso-asymmetry lines in the figures correspond to 
$\AsymRNCM = -2\%$, $-1\%$, $+1\%$,
$+2.5\%$, $+10\%$, $+20\%$,  $+40\%$, $+60\%$, $+80\%$, $+90\%$, $+100\%$.
The iso-(\daynight) asymmetry contours for the 
\night\ and \core\ samples shown in 
Figs. \FigContour a and \FigContour b
have been obtained both for $Y_e = 0.467$ (solid lines) and for
$Y_e = 0.5$ (dash-dotted lines).
Since the Earth effect depends on the 
quantity $Y_e \rho$, the case of $Y_e = 0.500$ and a given matter density
distribution in the
core is equivalent to the case of $Y_e = 0.467$ and a core matter density
which is uniformly increased by 7\%.  
For given \dms\ and \SdTvS\ the differences 
between \AsymRN\ or \AsymRM\ 
and \AsymRC\ are largest in the region
of the nonadiabatic (NA) or small mixing angle solution.

 The magnitude of the \daynight\ asymmetry 
in the region of the large mixing angle adiabatic (AD) 
solution
does not change 
significantly with the change of the sample.
As we have already pointed out in ref. \cite{ArticleI}, although
the \core\ enhancement of the \daynight\ effect at large mixing angles   
is noticeable, it is not dramatic.
The asymmetry \AsymRN\ (\AsymRC) 
is predicted to vary approximately between
0.01 and 0.35 (0.42) in the
``conservative'' region of the AD solution.   
%of about $+10\%$\ for the \daynight\ 
% asymmetry in the case the AD solution is the right one.
Since the neutrino transitions in the mantle 
are the dominant source of the Earth effect in this case, 
the same conclusion should be valid even if the data are 
taken over a period which represents a fraction of the year.

   In contrast, in the NA solution region the \daynight\ asymmetry 
in the \core\ sample
is enhanced by a factor of up to six. The reduction in 
statistics for this sample  
increases the statistical error by a factor of 2.65. Therefore the effective
enhancement is by about a factor of 2.3, which is quite significant.
% The iso - \daynight\ asymmetry contours  
% for the \core\ sample shown in Fig. 
% \FigContour b cross a large part of the ``conservative'' NA solution region.
The \daynight\ asymmetry   
for the \core\ sample is greater than 0.01 in absolute value in most of the
``conservative'' NA solution region (see Fig. \FigContour b).  
In the NA solution region derived in the reference solar model
\AsymRC\ takes values 
in the interval (0.025 - 0.20).
Our results show that the 
\AsymRC\ does not exceed approximately 0.30 in the region of the 
NA solution. It is interesting to note also 
that a negative Earth effect larger in absolute value than
1\%\ is predicted for the core sample for most of the  
values of the parameters from the ``conservative'' NA solution region.
Note that the +10\% effect line 
crosses only marginally the ``conservative'' NA solution region 
in the case of the \night\ sample 
(Fig. \FigContour a).
A comparison between Fig. \FigContour a  and  Fig. \FigContour b indicates 
that the \core\ selection is a very effective method of 
enhancing the \daynight\ 
asymmetry in the data sample despite the loss of statistics. 

  The difference between the values of the \core\ sample 
asymmetry \AsymRC\ calculated for $Y_e = 0.467$ and for
$Y_e = 0.5$ depends very strongly on the values of 
\dms\ and \SdTvS. 
For most of the values of \dms\ and \SdTvS\ from
the ``conservative'' regions of 
the two MSW solutions (see Fig. \FigContour b  and Table \TabAsymmetries)
this difference does not exceed 0.01 in absolute value. However, for certain 
\SdTvS\ and specific values of \dms\ from the 
NA solution region it can be larger
than 0.02 in absolute value. As Table II illustrates,
for $\SdTvS = 0.006; 0.008; 0.01; 0.013$,
\AsymRC\ can change respectively by -0.022; -0.047; -0.03; 
+0.026 when \Ye\ is changed from 0.467 to 0.500. This change
in the region of the AD solution is typically smaller
than 0.02 in absolute value, except for $\SdTvS \cong 0.56$ 
when the change can be
as large as by (-0.059). The values of the D-N asymmetry for the 
\night\ sample calculated for the two indicated values of \Ye\
typically do not differ by more 0.01 in absolute value. The only exception
is the case of 
$\SdTvS \cong 0.56$ for which the difference can be 
considerably larger (can be equal to (-0.039), for instance).
  
    The \daynight\ asymmetry in the \mantle\ sample is smaller
than 10\% in the ``conservative'' NA solution region;
in the AD solution region it can be as large as 35\%. 

   Because of the  large differences between the core and the 
mantle structures, the 
\mantle\ and \core\ subsamples provide two independent measurements of 
the \daynight\ effect. These can be combined to constrain better 
the neutrino parameters \dms\ and \SdTvS, 
utilizing the full statistics of the \night\ sample. 
Since the core enhancement is larger at small mixing angles, the NA 
solution region can be more effectively constrained 
than the AD solution region by the 
combined use of the   
\daynight\ asymmetry values in the \core\ and \mantle\ samples. 
In the NA solution region, the iso-asymmetry contour 
lines for the \mantle\ and the \core\ samples 
cross each other nearly perpendicularly. Therefore if 
$\SdTvS~\gsim~ 8 \times 10^{-3}$\
and $3.5 \times 10^{-6} \mbox{ eV}^2~ \lsim~ \dms~ \lsim~ 
6 \times 10^{-6} \mbox{ eV}^2$\ 
it should be possible to 
reduce considerably the allowed NA region by just
combining the experimental results for \AsymRC\ and 
\AsymRM.
Outside this region the \daynight\ effect can be detectable only in the 
\core\ sample for $\SdTvS~ \gsim ~5 \times 10^{-3}$.
If, for instance, a +10\% \daynight\ asymmetry will be detected 
in the \core\ sample
and no \daynight\ asymmetry will be observed in the 
\mantle\ sample, the NA solution 
region would be restricted to a narrow band along the $+10\%$ line,
limited approximately by 
$6\times 10^{-3}~ \lsim ~\SdTvS~ \lsim~ 8 \times 10^{-3}$\ 
and
$5\times 10^{-6} \mbox{ eV}^2 ~\lsim~ \dms~ \lsim~ 10^{-5} \mbox{ eV}^2$.
For $\SdTvS~ \lsim ~5 \times 10^{-3}$ the Earth 
effect is negative and it does not exceed 3.5\% in absolute value. 
A negative \daynight\ asymmetry larger than 0.01 in absolute value
is possible 
both for the \core\ and the \mantle\ samples in a small region 
of values of the parameters centered around 
$\SdTvS \approx 3 \times 10^{-3}$\ and $\dms \approx 3 \times 10^{-6}$\
eV$^2$.
% but this region is excluded both by the classical MSW analysis and by
% the extended one. 
For the allowed values of  
$\dms$ and $\SdTvS$ from the NA region derived within the reference 
solar model \cite{BP95}, the Earth effect is positive both 
for the core and the mantle samples.
However, \AsymRC\ can be negative for values of 
\dms and \SdTvS\ from the ``conservative'' NA solution region.
If a negative Earth effect with 
$|\AsymRC| \geq  0.02$\ will be detected, the NA solution
will be restricted to the narrow region determined by the inequalities  
$1.3 \times 10^{-3}~\lsim~ \SdTvS~ \lsim~ 3 \times 10^{-3}$\ 
and
$5 \times 10^{-6} \mbox{ eV}^2~\lsim~ \dms~ \lsim~ 7.5 \times 10^{-6} \mbox{ 
eV}^2$.

 Using the mean event rate solar neutrino data provided by 
the different solar neutrino detectors
and the measurements of the \daynight\ asymmetry in the Super-Kamiokande
experiment can allow to obtain information
not only about the parameters \dms\ and \SdTvS, but also 
about the value of the \BHt\ neutrino flux 
provided the solar neutrinos
undergo MSW transitions in the Sun and in the Earth. 
Indeed, a change of $\Phi_B$ shifts the region of the NA solution,
obtained for a given value of $\Phi_B$,
along the \SdTvS\ axis, leaving the region's shape and dimensions
essentially unchanged
(to less extent a similar behavior is exhibited
by the large mixing angle solution region).  
The iso - (\daynight) asymmetry contours cross nearly vertically
the ``conservative'' NA solution region, derived 
by varying $\Phi_B$ and using the predictions of the 
reference solar model \cite{BP95} for the other solar neutrino flux
components.  
Thus, the measurement of 
the \daynight\ asymmetry can allow to 
select an MSW solution of the solar neutrino problem
which corresponds to a given value of $\Phi_B$. 
For example, a $+10\%$\ 
\daynight\ effect in the NA region 
would correspond to a value of $\Phi_B$ expected in the reference
solar model \cite{BP95},
while the absence of a positive \daynight\ effect at $1\% - 2\%$\ level,
would indicate either that the MSW effect does not take place or that 
the \BHt\ neutrino flux is smaller than the flux in the 
reference model \cite{BP95}.
The \daynight\ asymmetry  
is independent on the \BHt\ neutrino flux, and the MSW probability
$\PTots(\nue \rightarrow \nue)$\ varies very little when calculated
in different standard solar models which are compatible with
the existing observational constraints
\footnote{The solar $\nu_e$ survival probability depends in the case 
of interest, as is well
known, on the electron number density distribution in the Sun.
All contemporary solar models compatible with the helioseismological
observations predict very similar electron number density
distributions in the Sun.}. 
Thus, the measurement of the \daynight\ asymmetry allows 
to determine \SdTvS\  and \dms\
essentially in a solar model independent way. 
The uncertainties in the iso - (\daynight) asymmetry contours 
associated with the  
Earth model being used in the calculations
are smaller than the uncertainties 
in the MSW solution regions
associated with the spread in the solar model predictions
for the value of the \BHt\ neutrino flux.

  We have studied also the dependence of the asymmetries
$\AsymRs(\TeTh)$\ for $\Ye = 0.467$ on the value of the
threshold energy \TeTh\ being used in the event selection,
by performing a calculation of \AsymRs\ 
for $\TeTh = 7.5$~MeV as well. The results of this analysis
are presented in Table III. We have found that in the 
cases of the \night\ and \mantle\ samples one can have
$|\AsymRNM(7.5~MeV) - \AsymRNM(5.0~MeV)| \geq 0.01$
only for $\SdTvS \gsim 0.013$, where $\AsymRs(7.5 MeV)$\ and
$\AsymRs(5.0 MeV)$\ are the values of the asymmetries for $\TeTh = 7.5$~MeV and
$\TeTh = 5.0~\MeV$. Moreover, when the indicated difference is larger than
0.01, it is always positive, i.e., we have  
$\AsymRNM(7.5~\MeV) > \AsymRNM(5.0~\MeV)$. This is due mainly 
to the fact that the event rate \ReD\ decreases with the increase
of \TeTh. It should be emphasized, however, that for the values
of $\SdTvS \gsim 0.013$\ and \dms\ from Table III, for which the 
calculations with $\TeTh = 7.5~\MeV$\ have been done, one has 
typically $|\AsymRNM(7.5~\MeV) - \AsymRNM(5.0~\MeV)| 
\sim (0.01 - 0.02)$.
The difference between the $\TeTh = 7.5~\MeV$ and the $\TeTh = 5.0~\MeV$\
asymmetries for the \night\ (\mantle) sample reaches the maximal value
of 0.051 (0.038) for $\SdTvS \cong 0.56$ 
and $\dms \cong 10^{-5}~eV^2$\ ($f_B = 1.5$). 
 
  As it follows form Table III, the dependence of the 
\core\ sample asymmetry \AsymRC\ on \TeTh\
is much stronger and more complicated. For
$sin^2 2\theta_V \geq 0.006$ one typically has
$|\AsymRC(7.5~\MeV) - \AsymRC(5.0~\MeV)| \geq 0.02$. This 
difference can take the values of -0.046, -0.076, -0.117 for 
$\SdTvS \cong 0.008,~0.01,~ 0.013$, i.e., the change of $T_{e,th}$
from 5 MeV to 7.5 MeV can decrease considerably the asymmetry \AsymRC.
This is not a general rule, however: for $\SdTvS \cong 0.008$, for
instance, \AsymRC\ increases by 0.034 when \TeTh\ is raised to 7.5 MeV if 
$\dms \cong 10^{-5}~eV^2$, and decreases by 0.046 if
$\dms \cong 5\times 10^{-6}~eV^2$. The indicated 
dependence can be exploited to obtain better constraints on the two 
neutrino transition parameters \dms\ and \SdTvS\ 
as well as on the value of the \BHt\ neutrino flux if
the existence of the D-N effect will be established.    
  
\section{Conclusions}

   Assuming that the solar neutrinos undergo two-neutrino MSW 
$\nu_e \rightarrow \nu_{\mu (\tau)}$ transitions which are at the origin of 
the solar neutrino problem and using the results of a high precision calculation
of the MSW solar $\nu_e$ survival probability performed in ref. \cite{ArticleI},
we have derived in the present article detailed predictions for several
D-N effect related observables for the Super-Kamiokande detector. The observables 
we have studied here are: i) the shape of the recoil-e$^{-}$ spectrum 
$\Ses(\Te)$, and the ratio $\deltaSs(\Te) = \Ses(\Te)/\SeZr(\Te)$,
$\SeZr(\Te)$ being the standard spectrum in the case of ``conventionally''
behaving (on the way to the Earth and the detector) solar \BHt\ neutrinos,   
ii) the D-N asymmetry in the spectrum, \AsymSs(\Te), and
iii) the energy-integrated D-N asymmetry, \AsymRs. The observables 
\Ses(\Te)\ (\deltaSs(\Te)), \AsymSs(\Te)\ and \AsymRs\
have been calculated for three different samples of events produced
respectively by solar neutrinos which cross the Earth mantle only ($s = M$),
the Earth core ($s = C$) and the mantle only + the core (full night, $s = N$).
The e$^{-}-$spectrum has been derived also for the 
sample of events detected during the day ($s = D$). Obviously,
\SeD(\Te)\ (\deltaSD(\Te)) is sensitive to the MSW effect in the Sun, while
\Ses(\Te)\ (\deltaSs(\Te)), \AsymSs(\Te)\ and \AsymRs, $s = N,C,M$,
depend both on the solar neutrino transitions in the Sun and in the Earth.

  All quantities for which we have obtained predictions, 
\Ses(\Te)\ (\deltaSs(\Te)), 
\AsymSs(\Te)\ and \AsymRs, are one year averaged. The energy-integrated
asymmetries \AsymRs\ have been calculated for e$^{-}$ kinetic energy threshold
of $T_{e,th} = 5~MeV$. The calculation of the \mantle\ (\core) sample observables 
has been performed for $Y_e = 0.50~(0.467)$.  We have studied the dependence
of the predictions for the \core\ sample observables on the uncertainties 
in the chemical composition and in the value of the bulk matter density
of the Earth core. The dependence of the asymmetries \AsymRN,
\AsymRC\ and \AsymRM\ on the 
threshold energy $T_{e,th}$ being used for the event selection 
was also investigated by comparing the values of the asymmetries
calculated for $T_{e,th} = 5.0~MeV$ and for $T_{e,th} = 7.5~MeV$ (Table III).
Iso - (D-N) asymmetry contour plots for the \night, \core\ and \mantle\
samples have been obtained for $\SdTvS \gsim 10^{-4}$, i.e., in
the ``conservative'' region of the MSW solution of the solar neutrino
problem, derived by varying the \BHt\ and \BeSv\ neutrino fluxes (see
refs. \cite{SPnu96,PKAS94}). 

    The results of the present study can be summarized as follows.

1. The \core\ sample D-N asymmetry \AsymRC\ can be strongly enhanced with respect
to the \night\ and \mantle\ sample asymmetries \AsymRN\ and \AsymRM.
Such an enhancement takes place in the region of the NA solution of the solar
neutrino problem, i.e., for $\SdTvS \leq 0.013$: it can be as large
as by a factor of six. In the region of the AD solution \AsymRC\ is larger
than \AsymRN\ by a factor which does not exceed 1.5.

2. The asymmetry \AsymRC\ is greater than 1\% in absolute value in most of the
``conservative'' NA solution region; it does not exceed 30\% in this region
(Fig. 3b). For certain \dms\ and \SdTvS\
we have \AsymRC$~ = - (2.0 - 3.5)$\%. In the ``conservative'' 
region of the AD solution \AsymRC\ is predicted to lie in the interval
(1.0 - 42.0)\%. 

   The \night\ sample asymmetry \AsymRN\ takes values between 1\% and 10\%
(1\% and 35\%) in the NA (AD) solution region (Fig. 3a). One has
\AsymRN$~ \geq 1.0$\% only in a small subregion of the ``conservative'' NA 
solution region (this subregion constitutes a part of the reference solar model
\cite{BP95} solution region). Similar conclusions are valid for the \mantle\
sample asymmetry \AsymRM\ (Fig. 3c).

3.  The difference between the values of \AsymRC\ 
calculated for $Y_e = 0.467$ and for
$Y_e = 0.50$ depends very strongly on the values of 
\dms\ and \SdTvS\ (Fig. 3b and Table II). 
Although for most \dms\ and \SdTvS\ from
the ``conservative'' MSW solution regions 
this difference does not exceed 0.01 in absolute value, for certain 
combinations of values of 
\SdTvS\ and \dms\ 
it can be as large as
-0.047 (-0.059) in the NA (AD) solution region.
The analogous difference for the \night\ sample asymmetry 
\AsymRN\ is typically smaller
than (0.01 - 0.02) in absolute value, one exception being
the case of $\SdTvS \sim 0.56$ when it can reach the value of
-0.039.

4. Raising the threshold energy $T_{e,th}$ can change 
 considerably the value of the
\core\ sample asymmetry \AsymRC: depending on the values of
\SdTvS\ and \dms\ it can either increase or
decrease \AsymRC (Table III). The dependence of 
\AsymRN\ and \AsymRM\ on $T_{e,th}$
is weaker. The sensitivity of \AsymRC\ to the value of 
$T_{e,th}$ can be used to  
better constrain the values of 
\SdTvS\ and \dms\ if 
the D-N effect will be observed. 
   
5. The measurement of the \core\ and \mantle\ sample asymmetries
\AsymRC\ and \AsymRM\ can be used not only to constrain
the allowed regions of values of the parameters \dms\ and
\SdTvS, but also to obtain information about the value of the
\BHt\ neutrino flux. 

6. The \core\ sample spectrum \SeC(\Te)\  
can differ significantly (both in shape and magnitude) from the \DAY\ sample
spectrum, \SeD(\Te). In certain cases the \core\ sample D-N asymmetry
in the spectrum \AsymSC(\Te)\ can reach values of (20 - 30)\% (Figs. 2.5,
2.6, 2.8) and even 50\% (Fig. 2.10).

  The above results indicate that the Super-Kamiokande experiment
might be able to test, in particular, the $\SdTvS \leq 0.01$ region
of the MSW solution of the solar neutrino problem by performing
selective D-N asymmetry measurements. In the case of observation of a 
nonzero D-N effect, these measurements can allow to determine the value of the
\BHt\ neutrino flux which is compatible with the hypothesis
that the solar $\nu_e$ undergo two-neutrino MSW 
$\nu_e \rightarrow \nu_{\mu (\tau)}$ transitions.

\section{Acknowledgments}
We are indebted to the ICARUS group of the University of Pavia  
and INFN, Sezione di Pavia, and especially to Prof. E. Calligarich, 
for allowing the use of their computing facilities for the present study.
S.T.P is grateful to A. Duda for useful discussions and 
to B. Machet and the other members of
L.P.T.H.E., Universit\'e de Paris 7, 
where part of the work for the present study has been done, for
the kind hospitality extended to him during his visit. 
M.M. wishes to thank Prof. A. Piazzoli
for constant interest in the work and support
and the International School for Advanced Studies, Trieste, 
where most of the work for this study has been done, 
for financial support.
We would like to thank also E. Lisi for informing us about the publication
\cite{Lisi:Montanino:1997} as well as for correspondence.
The work of S.T.P. was supported in part by the EEC grant ERBFMRXCT960090
and by Grant PH-510 from the Bulgarian Science Foundation.

%\newpage

%%%%%%%%%%%%%%%%
% Bibliography %
%%%%%%%%%%%%%%%%

\newpage

%%%%%%%%%%
% TABLES %
%%%%%%%%%%

%%%TABLE I

\begin{table}[h]
\bec
\begin{tabular}{|rccc||cccc|}
\multicolumn{8}{c}{\bf{Table \TabEventRates. Event Rates for the \SK\ Detector.}}\\ 
\hline
  \multicolumn{4}{|c||}{}   & \multicolumn{4}{|c|}{Event Rates $\Res/\ReZr$} \\ 
\multicolumn{1}{|c}{N.}   & \SdTvS  & \dms   & $f_B$ 
                            & \DAY     & \night & \core   & \mantle \\
\hline\hline
%N &  SdTvS  &   dms  &  F8 &   Day    &   Full &   Core  &  Mantle \\
 1 &  0.0008 &   9e-6 & 0.4 &  0.8441  & 0.8434 &  0.8396 &  0.8440 \\ 
 2 &  0.0008 &   7e-6 & 0.4 &  0.8742  & 0.8724 &  0.8633 &  0.8739 \\ 
 3 &  0.0008 &   5e-6 & 0.4 &  0.9052  & 0.9018 &  0.8936 &  0.9032 \\ 
\hline
 4 &  0.0010 &   9e-5 & 0.4 &  0.2413  & 0.2413 &  0.2413 &  0.2413 \\ 
 5 &  0.0010 &   7e-6 & 0.4 &  0.8472  & 0.8451 &  0.8347 &  0.8469 \\ 
 6 &  0.0010 &   5e-6 & 0.4 &  0.8847  & 0.8807 &  0.8713 &  0.8823 \\ 
\hline
 7 &  0.0020 &   1e-5 & 0.5 &  0.6426  & 0.6421 &  0.6400 &  0.6425 \\ 
 8 &  0.0020 &   7e-6 & 0.5 &  0.7271  & 0.7245 &  0.7119 &  0.7266 \\ 
 9 &  0.0020 &   5e-6 & 0.5 &  0.7905  & 0.7849 &  0.7726 &  0.7870 \\ 
\hline
10 &  0.0040 &   1e-5 & 1.0 &  0.4408  & 0.4414 &  0.4443 &  0.4410 \\ 
11 &  0.0040 &   7e-6 & 0.7 &  0.5469  & 0.5467 &  0.5463 &  0.5468 \\ 
12 &  0.0040 &   5e-6 & 0.7 &  0.6378  & 0.6341 &  0.6292 &  0.6349 \\ 
\hline
13 &  0.0060 &   1e-5 & 1.5 &  0.3233  & 0.3256 &  0.3365 &  0.3239 \\ 
14 &  0.0060 &   7e-6 & 1.0 &  0.4244  & 0.4289 &  0.4527 &  0.4251 \\ 
15 &  0.0060 &   5e-6 & 0.7 &  0.5224  & 0.5249 &  0.5380 &  0.5228 \\ 
\hline
16 &  0.0080 &   1e-5 & 1.5 &  0.2543  & 0.2584 &  0.2780 &  0.2552 \\ 
17 &  0.0080 &   7e-6 & 1.5 &  0.3406  & 0.3511 &  0.4038 &  0.3424 \\ 
18 &  0.0080 &   5e-6 & 1.0 &  0.4351  & 0.4464 &  0.4822 &  0.4404 \\ 
\hline
19 &  0.0100 &   7e-6 & 1.5 &  0.2830  & 0.2997 &  0.3823 &  0.2860 \\ 
20 &  0.0100 &   5e-6 & 1.0 &  0.3688  & 0.3901 &  0.4501 &  0.3802 \\ 
\hline
21 &  0.0130 &   5e-6 & 1.5 &  0.2979  & 0.3349 &  0.4302 &  0.3192 \\ 
\hline\hline
22 &  0.3000 & 1.5e-5 & 2.0 &  0.2171  & 0.2417 &  0.2479 &  0.2407 \\ 
23 &  0.3000 &   2e-5 & 2.0 &  0.2180  & 0.2354 &  0.2400 &  0.2346 \\ 
24 &  0.3000 &   3e-5 & 2.0 &  0.2214  & 0.2322 &  0.2341 &  0.2319 \\ 
25 &  0.3000 &   4e-5 & 2.0 &  0.2275  & 0.2351 &  0.2367 &  0.2349 \\ 
\hline
26 &  0.4800 &   3e-5 & 1.5 &  0.2717  & 0.2883 &  0.2909 &  0.2879 \\ 
27 &  0.4800 &   5e-5 & 1.5 &  0.2887  & 0.2975 &  0.2990 &  0.2972 \\ 
\hline
28 &  0.5000 &   2e-5 & 1.5 &  0.2735  & 0.3007 &  0.3053 &  0.3000 \\ 
\hline
29 &  0.5600 &   1e-5 & 1.5 &  0.2900  & 0.3678 &  0.4126 &  0.3604 \\ 
\hline
30 &  0.6000 &   8e-5 & 1.0 &  0.3685  & 0.3738 &  0.3746 &  0.3736 \\ 
\hline
31 &  0.7000 &   3e-5 & 1.0 &  0.3449  & 0.3682 &  0.3709 &  0.3677 \\ 
32 &  0.7000 &   5e-5 & 1.0 &  0.3588  & 0.3711 &  0.3731 &  0.3708 \\ 
\hline
33 &  0.7700 &   2e-5 & 1.0 &  0.3697  & 0.4082 &  0.4097 &  0.4079 \\ 
\hline
34 &  0.8000 & 1.3e-4 & 0.7 &  0.4744  & 0.4771 &  0.4777 &  0.4770 \\ 
\hline
35 &  0.9000 &   4e-5 & 0.7 &  0.4438  & 0.4644 &  0.4672 &  0.4639 \\ 
36 &  0.9000 &   1e-4 & 0.7 &  0.4765  & 0.4820 &  0.4829 &  0.4819 \\ 
\hline
\end{tabular}
\eec
\noindent
The results correspond to $\Ye = 0.467$. The event rates \Res\ and 
\ReZr\ (see eqs. (3) and (4)) were computed for $\TeTh = 5~$MeV. The 
factor $f_B$ is the ratio of the value of the \BHt\ neutrino flux used 
in the MSW analysis and the flux predicted in the reference solar 
model \cite{BP95}. The predicted event rates (in units of \ReZr) are
given by the product $f_{B}~\Res/\ReZr$.
\end{table}

\newpage

%%%TABLE II

\begin{table}[h]
\bec          
\begin{tabular}{|rccc||rrrc||rrrc|}
\multicolumn{12}{c}{{\bf{Table \TabAsymmetries.} D - N Asymmetries for 
the \SK\ Detector.}}\\ \hline
  \multicolumn{4}{|l||}{}&
  \multicolumn{4}{|c||}{$Y_e = 0.467$}&
  \multicolumn{4}{|c|}{$Y_e = 0.500$} \\
%%%%
  \multicolumn{4}{|l||}{}
  &
  \multicolumn{3}{|c}{$\AsymRs \times 100$}
  &
  \multicolumn{1}{c||}{$\frac{|\AsymRC|}{|\AsymRN|}$} 
  &
  \multicolumn{3}{|c}{$\AsymRs \times 100$}
  &
  \multicolumn{1}{c|}{$\frac{|\AsymRC|}{|\AsymRN|}$} 
  \\
%%%%
\multicolumn{1}{|c}{N.}   & \SdTvS  & \dms   & $f_B$ & 
%%%%
\multicolumn{1}{|c}{\night}   &
\multicolumn{1}{c}{\core}      &
\multicolumn{1}{c}{\mantle}    &
\multicolumn{1}{c||}{}   &
%\multicolumn{1}{c||}{$\frac{|\AsymRC|}{|\AsymRN|}$}   &
%%%%
\multicolumn{1}{|c}{\night}      &
\multicolumn{1}{c}{\core}      &
\multicolumn{1}{c}{\mantle}    &
\multicolumn{1}{c|}{}    \\
%\multicolumn{1}{c|}{$\frac{|\AsymRC|}{|\AsymRN|}$}    \\
    \hline\hline
%N &  SdTvS  &   dms  &  F8 &  AF   &    AC  &    AM  &  Ratio &   AF  &   AC  &   AM  & Ratio\\
1  & 0.0008 & 9.0e-6  & 0.4 & -0.09 &  -0.54 &  -0.01 &   6.24 & -0.12 & -0.75 & -0.01 &  6.25\\
2  & 0.0008 & 7.0e-6  & 0.4 & -0.21 &  -1.26 &  -0.04 &   6.00 & -0.22 & -1.35 & -0.04 &  6.14\\
3  & 0.0008 & 5.0e-6  & 0.4 & -0.37 &  -1.29 &  -0.23 &   3.44 & -0.35 & -1.14 & -0.23 &  3.26\\
\hline
4  & 0.0010 & 9.0e-5  & 0.4 &  3e-3 &   4e-3 &   3e-3 &   1.25 & 3e-3  &  4e-3 &  3e-3 &  1.33\\
5  & 0.0010 & 7.0e-6  & 0.4 & -0.25 &  -1.50 &  -0.04 &   6.01 & -0.26 & -1.59 & -0.04 &  6.12\\
6  & 0.0010 & 5.0e-6  & 0.4 & -0.45 &  -1.52 &  -0.27 &   3.39 & -0.43 & -1.35 & -0.27 &  3.14\\
\hline
7  & 0.0020 & 1.0e-5  & 0.5 & -0.07 &  -0.41 &  -0.01 &   6.00 & -0.10 & -0.62 & -0.01 &  6.20\\
8  & 0.0020 & 7.0e-6  & 0.5 & -0.35 &  -2.10 &  -0.07 &   5.98 & -0.36 & -2.18 & -0.07 &  6.06\\
9  & 0.0020 & 5.0e-6  & 0.5 & -0.71 &  -2.30 &  -0.45 &   3.23 & -0.67 & -2.02 & -0.45 &  3.01\\
\hline
10 & 0.0040 & 1.0e-5  & 1.0 &  0.15 &   0.79 &   0.04 &   5.42 & 0.22  &  1.28 &  0.04 &  5.82\\
11 & 0.0040 & 7.0e-6  & 0.7 & -0.04 &  -0.12 &  -0.02 &   3.48 & 0.01  &  0.21 & -0.02 & 21.00\\ 
12 & 0.0040 & 5.0e-6  & 0.7 & -0.59 &  -1.36 &  -0.46 &   2.31 & -0.56 & -1.12 & -0.47 &  2.00\\
\hline
13 & 0.0060 & 1.0e-5  & 1.5 &  0.72 &   3.98 &   0.17 &   5.56 &  1.05 &  6.20 &  0.17 &  5.90\\ 
14 & 0.0060 & 7.0e-6  & 1.0 &  1.06 &   6.46 &   0.17 &   6.13 &  1.26 &  7.60 &  0.17 &  6.03\\
15 & 0.0060 & 5.0e-6  & 0.7 &  0.47 &   2.93 &   0.06 &   6.19 &  0.46 &  2.82 &  0.06 &  6.13\\
\hline
16 & 0.0080 & 1.0e-5  & 1.5 &  1.63 &   8.93 &   0.37 &   5.47 &  2.36 & 13.59 &  0.37 &  5.76\\ 
17 & 0.0080 & 7.0e-6  & 1.5 &  3.04 &  17.00 &   0.53 &   5.59 &  3.39 & 19.10 &  0.53 &  5.63\\ 
18 & 0.0080 & 5.0e-6  & 1.0 &  2.55 &  10.26 &   1.22 &   4.02 &  2.44 &  9.54 &  1.22 &  3.91\\
\hline
19 & 0.0100 & 7.0e-6  & 1.5 &  5.72 &  29.85 &   1.05 &   5.22 &  6.28 & 32.85 &  1.05 &  5.23\\
20 & 0.0100 & 5.0e-6  & 1.0 &  5.60 &  19.84 &   3.03 &   3.54 &  5.36 & 18.38 &  3.03 &  3.43\\
\hline
21 & 0.0130 & 5.0e-6  & 1.5 & 11.69 &  36.33 &   6.89 &   3.11 & 11.21 & 33.72 &  6.89 &  3.01\\
\hline\hline
22 & 0.3000 & 1.5e-5  & 2.0 & 10.73 &  13.25 &  10.31 &   1.24 & 11.03 & 15.29 & 10.30 &  1.39\\
23 & 0.3000 & 2.0e-5  & 2.0 &  7.64 &   9.60 &   7.31 &   1.26 &  7.79 & 10.66 &  7.31 &  1.37\\
24 & 0.3000 & 3.0e-5  & 2.0 &  4.74 &   5.54 &   4.61 &   1.17 &  4.78 &  5.85 &  4.61 &  1.22\\
25 & 0.3000 & 4.0e-5  & 2.0 &  3.29 &   3.93 &   3.18 &   1.20 &  3.31 &  4.07 &  3.19 &  1.23\\
\hline
26 & 0.4800 & 3.0e-5  & 1.5 &  5.95 &   6.81 &   5.81 &   1.15 &  5.99 &  7.08 &  5.81 &  1.18\\
27 & 0.4800 & 5.0e-5  & 1.5 &  2.98 &   3.50 &   2.89 &   1.17 &  2.99 &  3.57 &  2.89 &  1.19\\
\hline
28 & 0.5000 & 2.0e-5  & 1.5 &  9.48 &  10.97 &   9.24 &   1.16 &  9.60 & 11.79 &  9.23 &  1.23\\
\hline
29 & 0.5600 & 1.0e-5  & 1.5 & 23.65 &  34.9  &  21.64 &   1.48 & 27.50 & 40.77 & 25.11 &  1.48\\
\hline
30 & 0.6000 & 8.0e-5  & 1.0 &  1.42 &   1.64 &   1.38 &   1.15 &  1.42 &  1.65 &  1.38 &  1.16\\
\hline
31 & 0.7000 & 3.0e-5  & 1.0 &  6.54 &   7.26 &   6.41 &   1.11 &  6.56 &  7.43 &  6.41 &  1.13\\
32 & 0.7000 & 5.0e-5  & 1.0 &  3.37 &   3.90 &   3.29 &   1.16 &  3.38 &  3.95 &  3.29 &  1.17\\
\hline
33 & 0.7700 & 2.0e-5  & 1.0 &  9.89 &  10.27 &   9.83 &   1.04 &  9.94 & 10.59 &  9.83 &  1.07\\
\hline
34 & 0.8000 & 1.3e-4  & 0.7 &  0.57 &   0.69 &   0.55 &   1.21 &  0.57 &  0.69 &  0.55 &  1.21\\
\hline
35 & 0.9000 & 4.0e-5  & 0.7 &  4.53 &   5.12 &   4.42 &   1.13 &  4.53 &  5.17 &  4.42 &  1.14\\
36 & 0.9000 & 1.0e-4  & 0.7 &  1.15 &   1.33 &   1.13 &   1.15 &  1.15 &  1.33 &  1.13 &  1.16\\
   \hline              
\end{tabular}            
\eec                     
\end{table}

%%% Table III

\begin{table}[h]
\bec          
\begin{tabular}{|rccc||rrrc||rrrc|}
\multicolumn{12}{c}{{\bf{Table \TabAsymmetriesMeV.} Threshold Energy Dependence
of the D - N Asymmetries ($Y_e = 0.467$).}}\\ \hline
  \multicolumn{4}{|l||}{}&
  \multicolumn{4}{|c||}{$\TeTh = 5$ MeV}&
  \multicolumn{4}{|c|}{$\TeTh = 7.5$ MeV} \\
%%%%
  \multicolumn{4}{|l||}{}
  &
  \multicolumn{3}{|c}{$\AsymRs \times 100$}
  &
  \multicolumn{1}{c||}{$\frac{|\AsymRC|}{|\AsymRN|}$} 
  &
  \multicolumn{3}{|c}{$\AsymRs \times 100$}
  &
  \multicolumn{1}{c|}{$\frac{|\AsymRC|}{|\AsymRN|}$} 
  \\
%%%%
\multicolumn{1}{|c}{N.}   & \SdTvS  & \dms   & $f_B$ & 
%%%%
\multicolumn{1}{|c}{\night}   &
\multicolumn{1}{c}{\core}      &
\multicolumn{1}{c}{\mantle}    &
\multicolumn{1}{c||}{}   &
%\multicolumn{1}{c||}{$\frac{|\AsymRC|}{|\AsymRN|}$}   &
%%%%
\multicolumn{1}{|c}{\night}      &
\multicolumn{1}{c}{\core}      &
\multicolumn{1}{c}{\mantle}    &
\multicolumn{1}{c|}{}    \\
%\multicolumn{1}{c|}{$\frac{|\AsymRC|}{|\AsymRN|}$}    \\
    \hline\hline
%N &  SdTvS  &   dms  &  F8 &  AF   &    AC  &    AM  &  Ratio &   AF  &   AC  &   AM  & Ratio\\
1  & 0.0008 & 9.0e-6  & 0.4 & -0.09 &  -0.54 &  -0.01 &   6.24 &  -0.13 &  -0.80 &  -0.02 &   6.30\\
2  & 0.0008 & 7.0e-6  & 0.4 & -0.21 &  -1.26 &  -0.04 &   6.00 &  -0.29 &  -1.70 &  -0.06 &   5.91\\
3  & 0.0008 & 5.0e-6  & 0.4 & -0.37 &  -1.29 &  -0.23 &   3.44 &  -0.43 &  -0.98 &  -0.34 &   2.30\\
\hline
4  & 0.0010 & 9.0e-5  & 0.4 &  3e-3 &   4e-3 &   3e-3 &   1.25 &   0.01 &   0.01 &   0.01 &   1.12\\
5  & 0.0010 & 7.0e-6  & 0.4 & -0.25 &  -1.50 &  -0.04 &   6.01 &  -0.34 &  -2.02 &  -0.07 &   5.90\\
6  & 0.0010 & 5.0e-6  & 0.4 & -0.45 &  -1.52 &  -0.27 &   3.39 &  -0.51 &  -1.18 &  -0.41 &   2.29\\
\hline
7  & 0.0020 & 1.0e-5  & 0.5 & -0.07 &  -0.41 &  -0.01 &   6.00 &  -0.11 &  -0.65 &  -0.02 &   6.10\\
8  & 0.0020 & 7.0e-6  & 0.5 & -0.35 &  -2.10 &  -0.07 &   5.98 &  -0.50 &  -2.92 &  -0.10 &   5.87\\
9  & 0.0020 & 5.0e-6  & 0.5 & -0.71 &  -2.30 &  -0.45 &   3.23 &  -0.84 &  -1.86 &  -0.67 &   2.21\\
\hline
10 & 0.0040 & 1.0e-5  & 1.0 &  0.15 &   0.79 &   0.04 &   5.42 &   0.17 &   1.02 &   0.03 &   5.89\\
11 & 0.0040 & 7.0e-6  & 0.7 & -0.04 &  -0.12 &  -0.02 &   3.48 &  -0.16 &  -0.79 &  -0.06 &   4.93\\
12 & 0.0040 & 5.0e-6  & 0.7 & -0.59 &  -1.36 &  -0.46 &   2.31 &  -0.85 &  -1.61 &  -0.73 &   1.89\\
\hline
13 & 0.0060 & 1.0e-5  & 1.5 &  0.72 &   3.98 &   0.17 &   5.56 &   0.93 &   5.43 &   0.17 &   5.84\\
14 & 0.0060 & 7.0e-6  & 1.0 &  1.06 &   6.46 &   0.17 &   6.13 &   1.14 &   6.78 &   0.18 &   5.93\\
15 & 0.0060 & 5.0e-6  & 0.7 &  0.47 &   2.93 &   0.06 &   6.19 &   0.10 &   0.91 &  -0.04 &   9.15\\
\hline
16 & 0.0080 & 1.0e-5  & 1.5 &  1.63 &   8.93 &   0.37 &   5.47 &   2.17 &  12.30 &   0.39 &   5.66\\
17 & 0.0080 & 7.0e-6  & 1.5 &  3.04 &  17.00 &   0.53 &   5.59 &   3.44 &  18.87 &   0.64 &   5.48\\
18 & 0.0080 & 5.0e-6  & 1.0 &  2.55 &  10.26 &   1.22 &   4.02 &   2.08 &   5.62 &   1.48 &   2.70\\
\hline
19 & 0.0100 & 7.0e-6  & 1.5 &  5.72 &  29.85 &   1.05 &   5.22 &   6.65 &  33.61 &   1.33 &   5.06\\
20 & 0.0100 & 5.0e-6  & 1.0 &  5.60 &  19.84 &   3.03 &   3.54 &   5.09 &  12.23 &   3.86 &   2.40\\
\hline
21 & 0.0130 & 5.0e-6  & 1.5 & 11.69 &  36.33 &   6.89 &   3.11 &  11.34 &  24.62 &   8.95 &   2.17\\
\hline\hline
22 & 0.3000 & 1.5e-5  & 2.0 & 10.73 &  13.25 &  10.31 &   1.24 &  12.61 &  15.40 &  12.14 &   1.22\\
23 & 0.3000 & 2.0e-5  & 2.0 &  7.64 &   9.60 &   7.31 &   1.26 &   9.03 &  11.99 &   8.53 &   1.33\\
24 & 0.3000 & 3.0e-5  & 2.0 &  4.74 &   5.54 &   4.61 &   1.17 &   5.57 &   6.45 &   5.42 &   1.16\\
25 & 0.3000 & 4.0e-5  & 2.0 &  3.29 &   3.93 &   3.18 &   1.20 &   3.99 &   4.78 &   3.86 &   1.20\\
\hline
26 & 0.4800 & 3.0e-5  & 1.5 &  5.95 &   6.81 &   5.81 &   1.15 &   6.95 &   7.85 &   6.80 &   1.13\\
27 & 0.4800 & 5.0e-5  & 1.5 &  2.98 &   3.50 &   2.89 &   1.17 &   3.64 &   4.27 &   3.53 &   1.17\\
\hline
28 & 0.5000 & 2.0e-5  & 1.5 &  9.48 &  10.97 &   9.24 &   1.16 &  11.07 &  13.36 &  10.69 &   1.21\\
\hline
29 & 0.5600 & 1.0e-5  & 1.5 & 23.65 &  34.9  &  21.64 &   1.48 &  28.78 &  46.48 &  25.45 &   1.62\\
\hline
30 & 0.6000 & 8.0e-5  & 1.0 &  1.42 &   1.64 &   1.38 &   1.15 &   1.85 &   2.13 &   1.81 &   1.15\\
\hline
31 & 0.7000 & 3.0e-5  & 1.0 &  6.54 &   7.26 &   6.41 &   1.11 &   7.56 &   8.26 &   7.44 &   1.09\\
32 & 0.7000 & 5.0e-5  & 1.0 &  3.37 &   3.90 &   3.29 &   1.16 &   4.05 &   4.67 &   3.95 &   1.15\\
\hline
33 & 0.7700 & 2.0e-5  & 1.0 &  9.89 &  10.27 &   9.83 &   1.04 &  11.34 &  11.91 &  11.25 &   1.05\\
\hline
34 & 0.8000 & 1.3e-4  & 0.7 &  0.57 &   0.69 &   0.55 &   1.21 &   0.78 &   0.93 &   0.75 &   1.20\\
\hline
35 & 0.9000 & 4.0e-5  & 0.7 &  4.53 &   5.12 &   4.42 &   1.13 &   5.32 &   6.02 &   5.20 &   1.13\\
36 & 0.9000 & 1.0e-4  & 0.7 &  1.15 &   1.33 &   1.13 &   1.15 &   1.46 &   1.63 &   1.43 &   1.12\\
   \hline              
\end{tabular}            
\eec                     
\end{table}

%%%%%%%%%%%
% FIGURES %
%%%%%%%%%%%

\newpage

\bec{\Large\bf{Figure Captions}}\eec

\noindent
{\bf{Figure \FigSKSpectra.}}
Recoil - e$^{-}$ spectrum to be measured in the \SK\ experiment.
Shown in the figure are the standard spectrum
$\SeZr(\Te)$\ (upper solid line) and the one year average
\DAY\ (long-dashed line), \night\ (short-dashed line), \core\
(lower solid line) and \mantle\ (dotted line) sample spectra 
in the case of a matter effect with parameters
 $\SdTvS = 0.01$ and
$\dms = 5 \times 10^{-5}$\ eV$^2$\ 
and for $Y_e = 0.467$. All spectra are normalized 
to the event rate in the \SK detector, predicted in 
the reference solar model \cite{BP95}\ for $\TeTh = 5$ MeV 

\vspace{1cm}

\noindent
{\bf{Figures 2.1 - 2.12.}}
Recoil - e$^{-}$ spectrum distortion $\deltaSs(\Te)$ (upper frame) 
and D-N asymmetry in the spectrum \AsymSs\ (lower frame) (see eqs. (4) and 
(5)) for the \DAY\ (long-dashed line), \night\ (short-dashed line),
\core\ (solid line) and \mantle\ (dotted line) samples
for $Y_e = 0.467$. The values of \dms\ and \SdTvS\
are indicated between the upper and the lower frames.

\vspace{1cm}
\noindent
{\bf{Figures 3a - 3c.}}
Iso - (D-N) asymmetry contour plots for the one year average 
\night\ (a), \core\ (b) and \mantle\ (c) samples fot the 
\SK\ detector. The solid (dash-dotted) lines in figures 
a and b correspond to
$\Ye = 0.467~(0.500)$. The dashed lines show the ``conservative''
MSW solution regions, while the dotted lines indicate the MSW 
solution regions obtained within the reference solar
model \cite{BP95}.

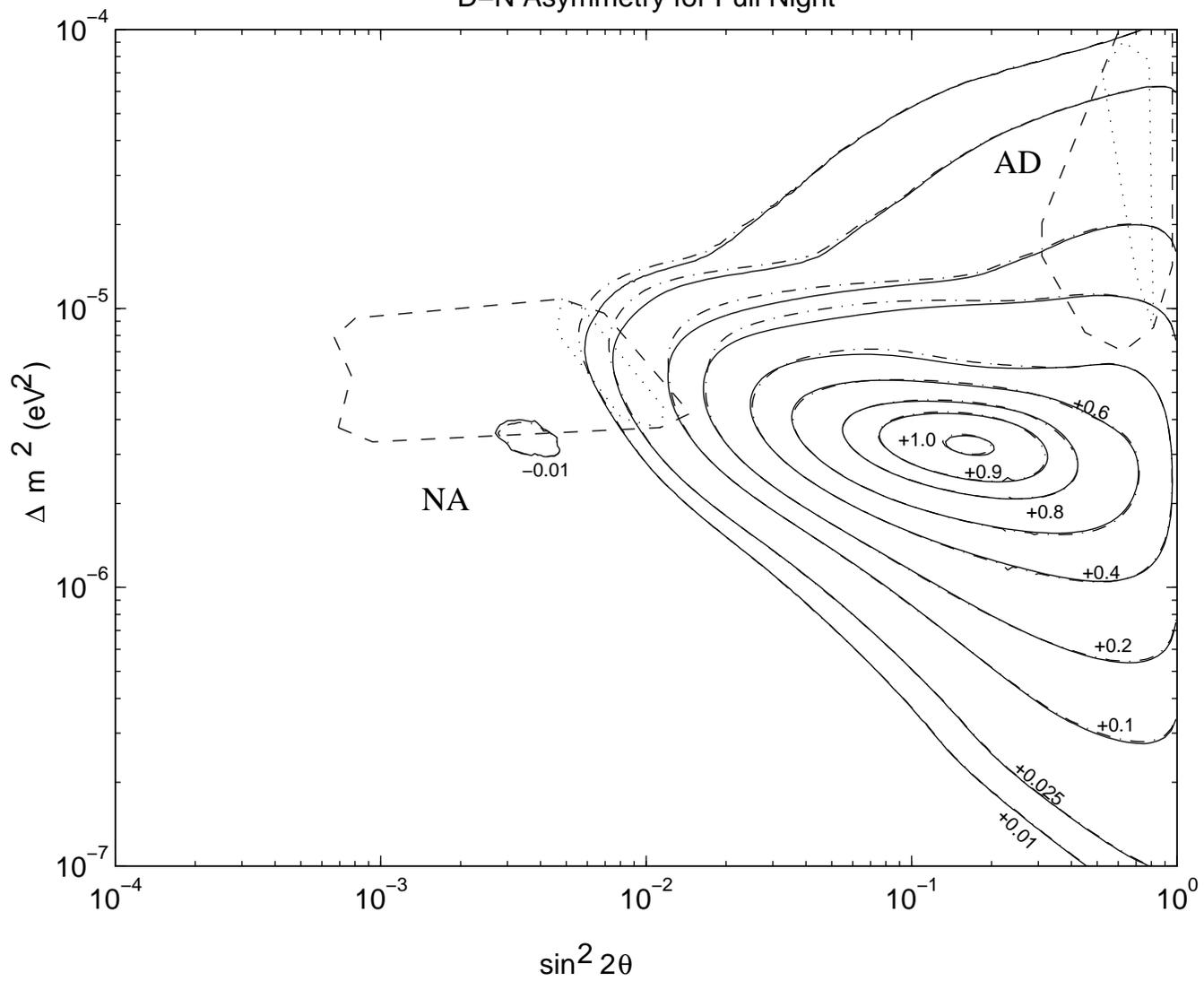
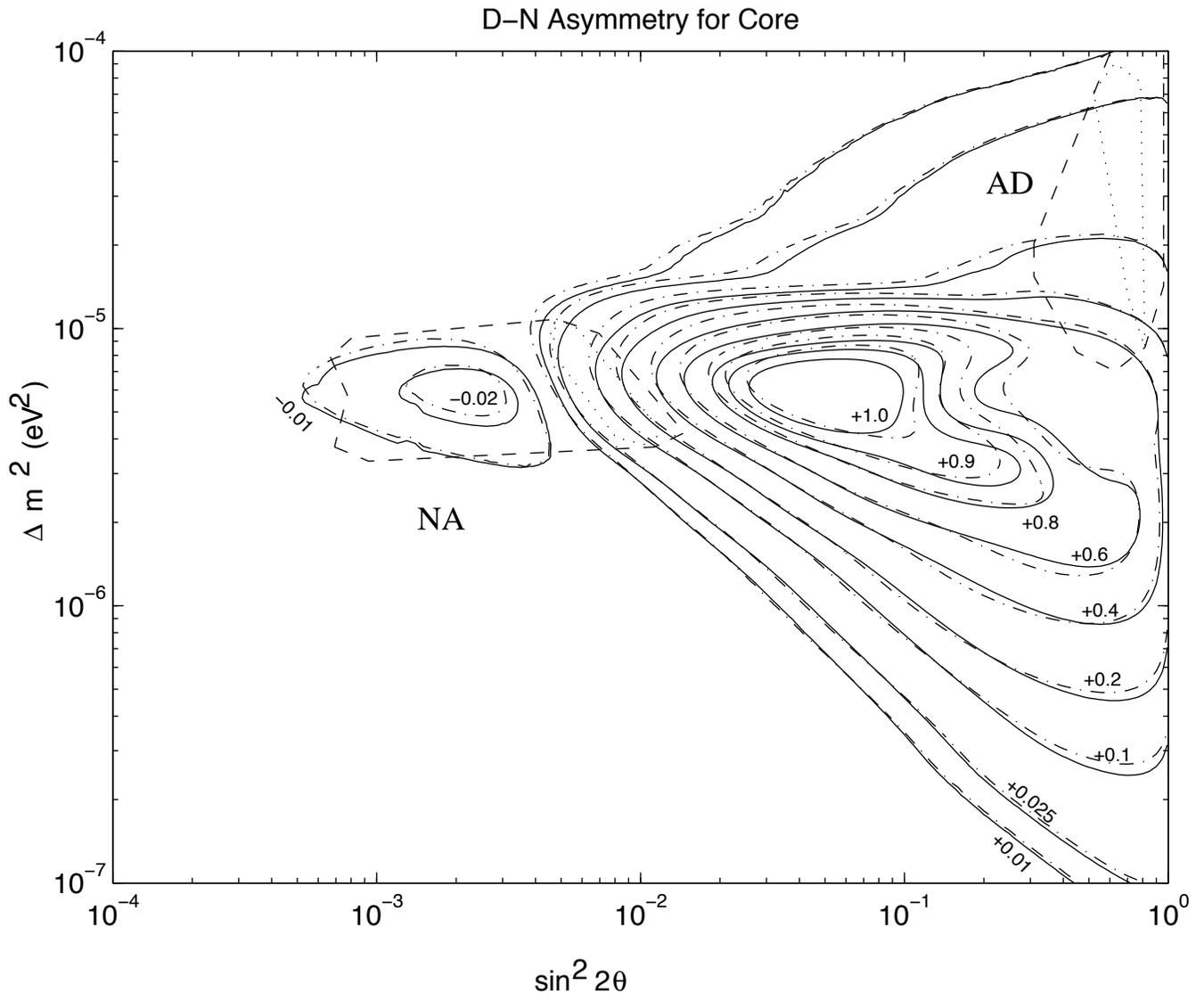
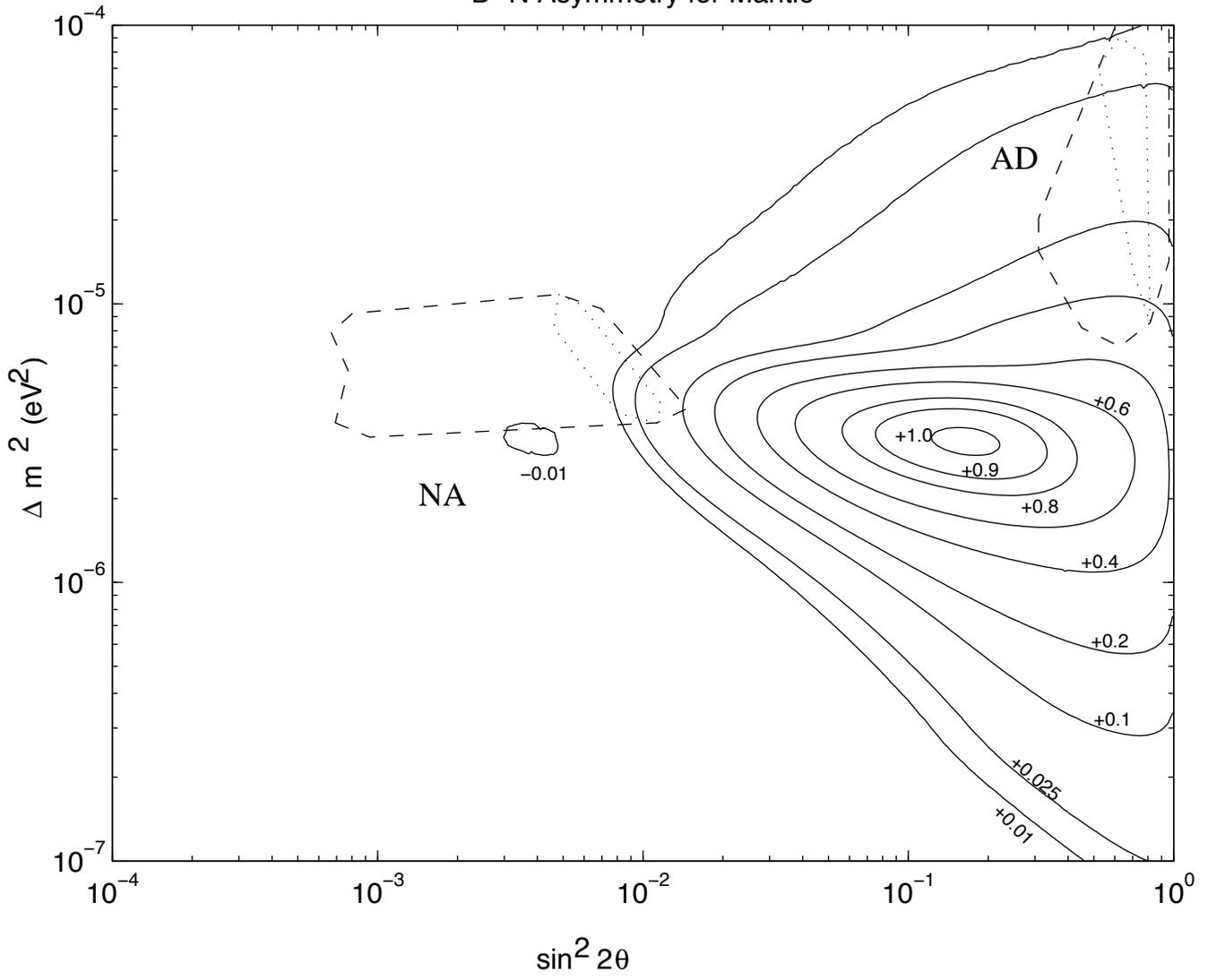
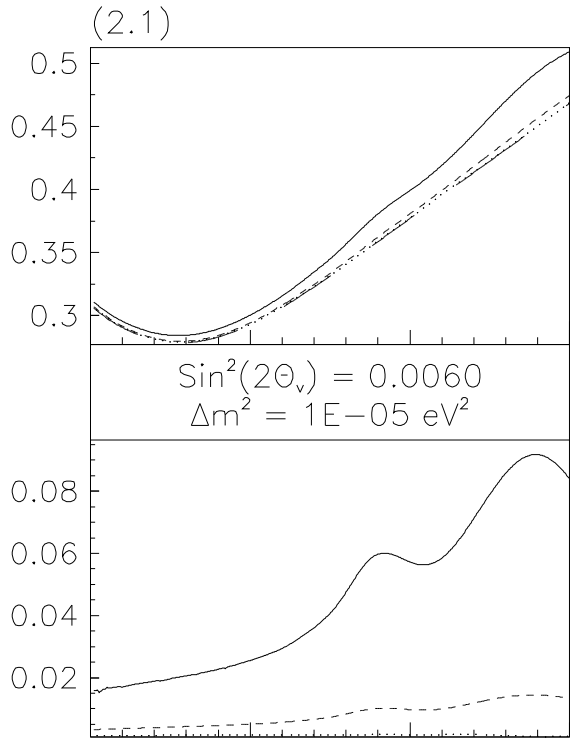
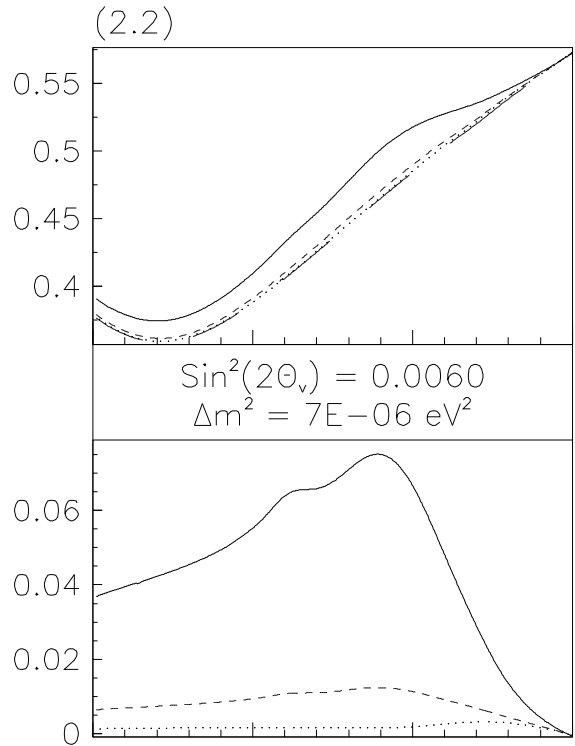
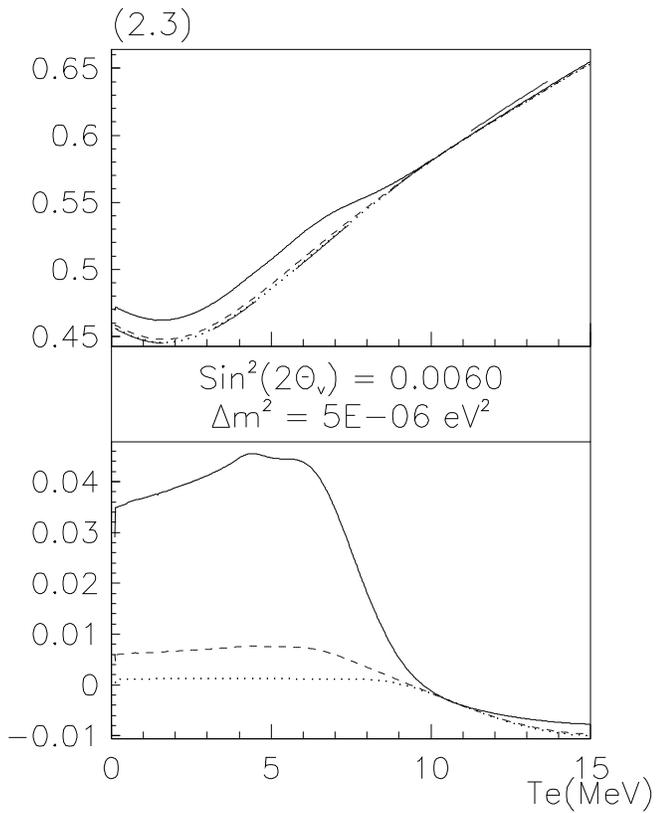
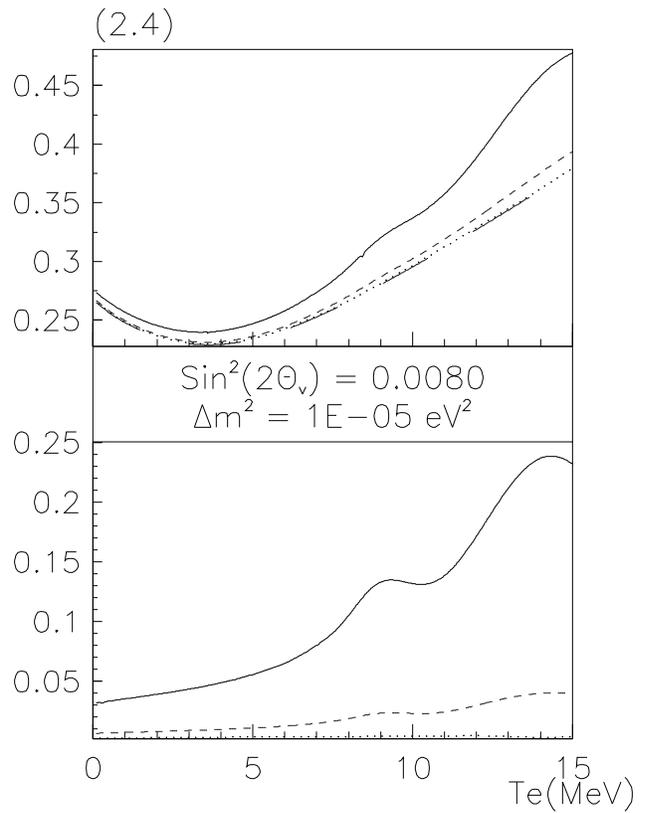
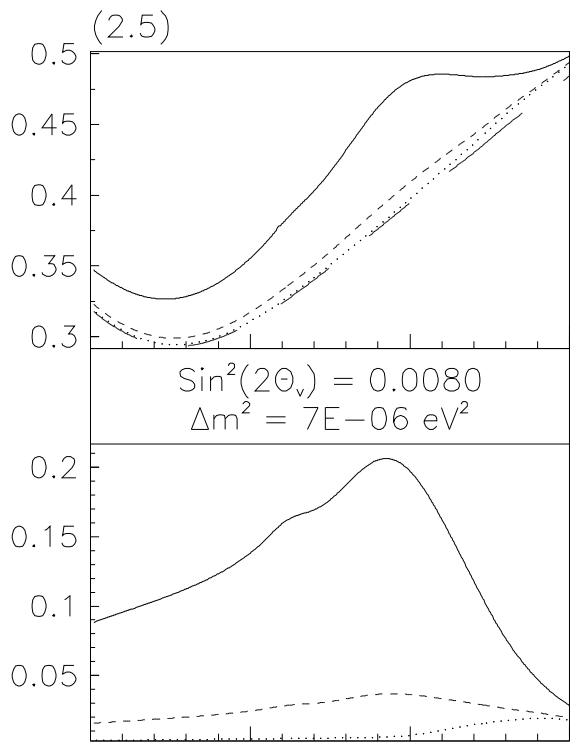
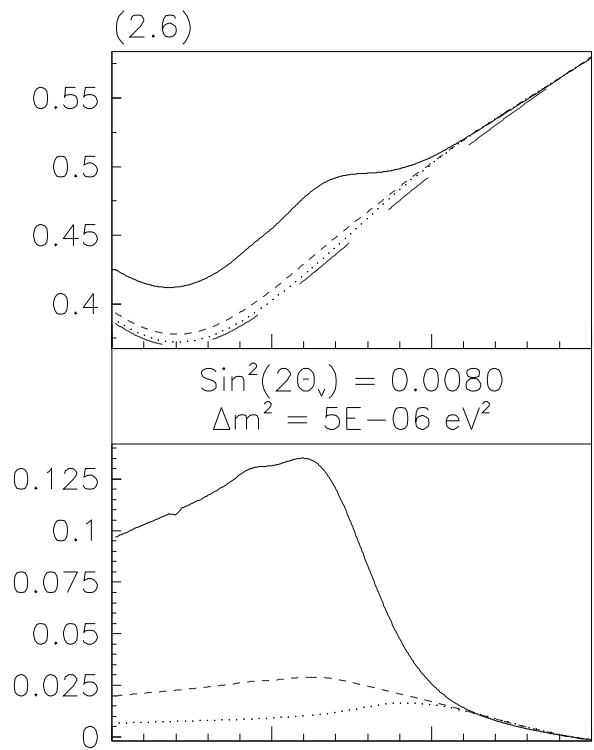
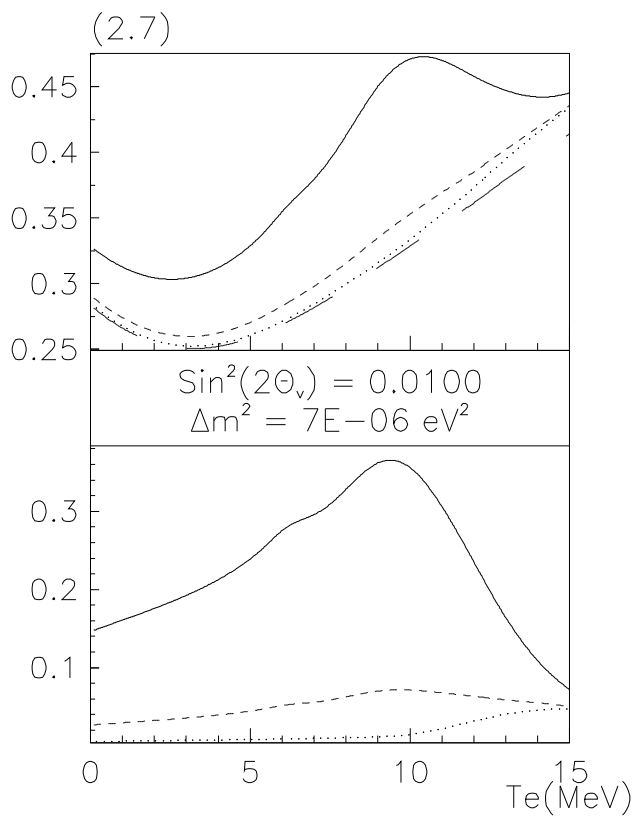
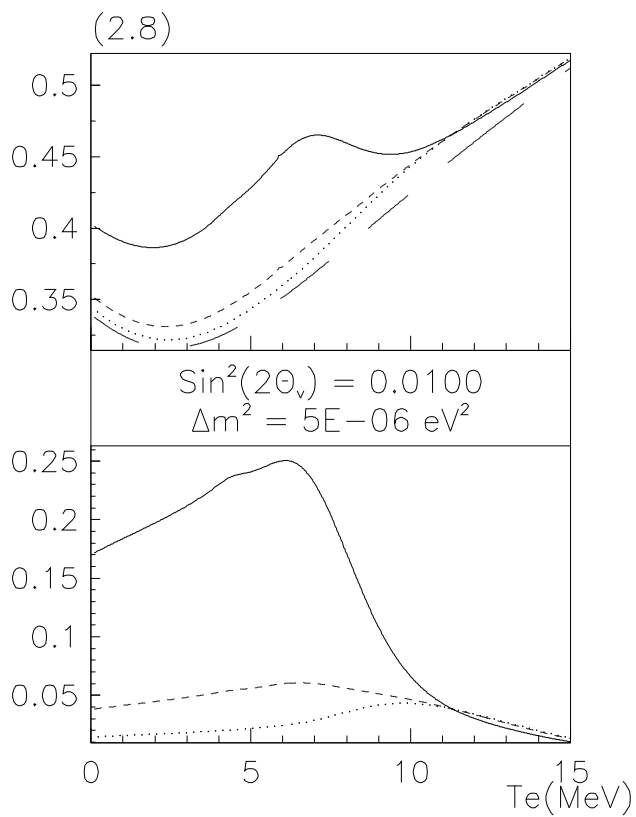
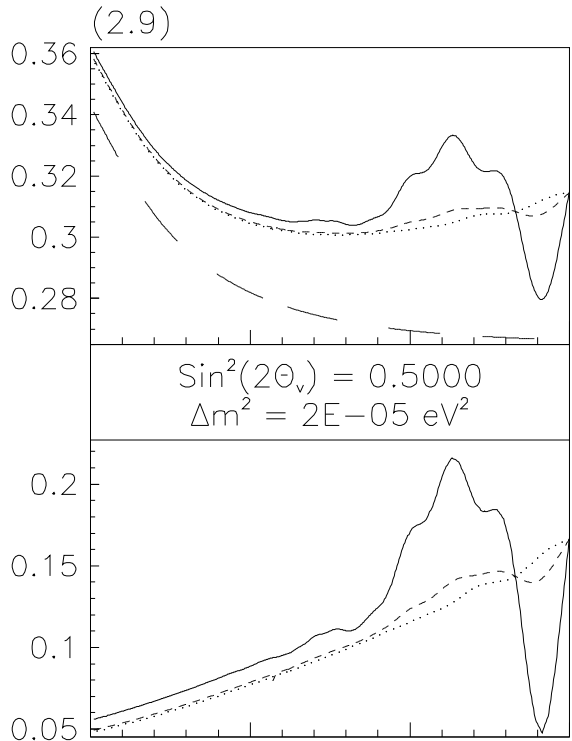
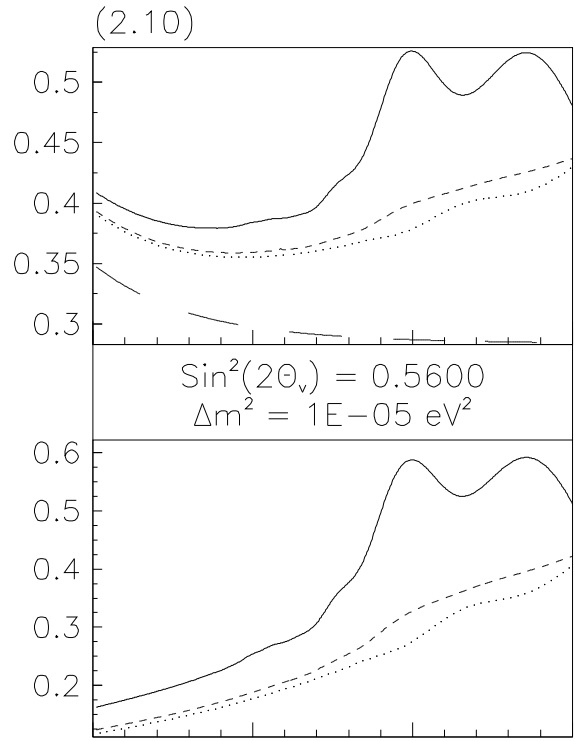
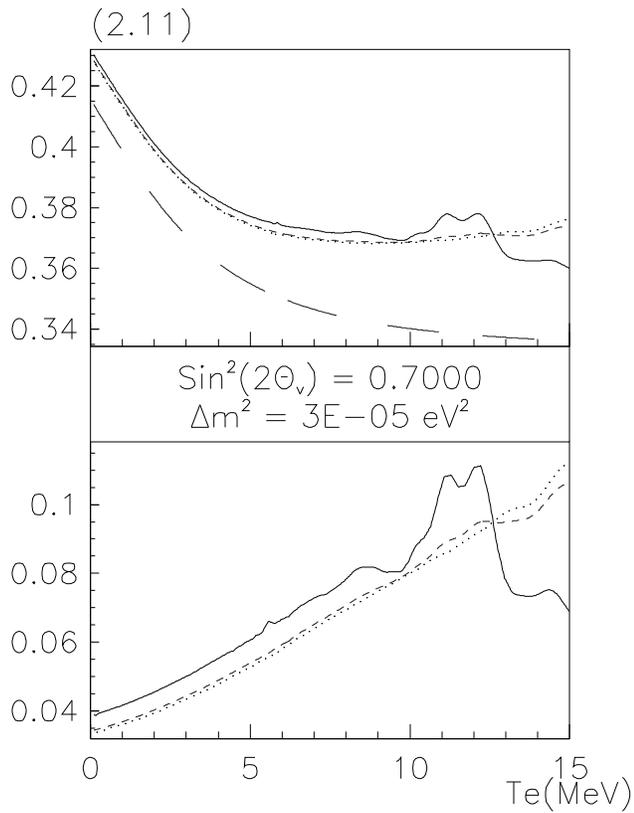
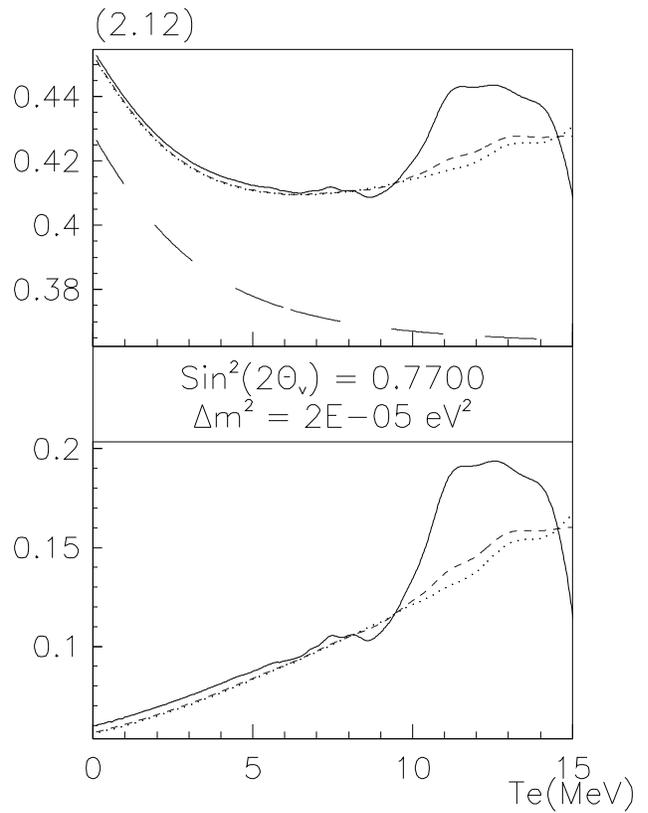
\end{document}